\documentclass[aps,twocolumn,prb,notitlepage,superscriptaddress]{revtex4-2}
\usepackage{amsmath,amssymb}
\usepackage{dsfont}
\usepackage{graphicx}
\usepackage{hyperref}
\usepackage{physics}
\usepackage{amsmath,amssymb}
\usepackage{dsfont}
\usepackage{mathtools}
\usepackage{cancel}
\usepackage{bbold}
\usepackage{bm}
\usepackage{color} 
\usepackage[dvipsnames]{xcolor}

\usepackage[caption=false]{subfig}

\usepackage{tcolorbox}

\usepackage{soul}

\newcommand{\beq}{\begin{equation}}
\newcommand{\eeq}{\end{equation}}

\newcommand{\EVS}{E_{\mathrm{V}}}

\newcommand{\DEZ}[1]{\Delta E_{Z}^{#1}}
\newcommand{\AEZ}[1]{\bar{E}_{Z}^{#1}}
\newcommand{\EZ}[1]{E_{Z}^{#1}}
\newcommand{\EVD}[1]{E_{V}^{#1}}

\newcommand{\du}{\downarrow\uparrow}
\newcommand{\ud}{\uparrow\downarrow}

\newcommand{\meanqs}[1]{\overline{#1}}

\newcommand{\optional}[1]{}  

\begin{document}
\author{\L{}ukasz Cywi{\'n}ski}\email{lcyw@ifpan.edu.pl}
\affiliation{Institute of Physics, Polish Academy of Sciences, al.~Lotnik{\'o}w 32/46, PL 02-668 Warsaw, Poland}
\author{Mats Volmer}
\affiliation{JARA-FIT Institute for Quantum Information, Forschungszentrum J\"ulich GmbH and RWTH Aachen University, Aachen, Germany}
\author{Tom Struck}
\affiliation{JARA-FIT Institute for Quantum Information, Forschungszentrum J\"ulich GmbH and RWTH Aachen University, Aachen, Germany}
\author{Giordano Scappucci}
\affiliation{QuTech and Kavli Institute of Nanoscience, Delft University \\ of Technology, Lorentzweg 1, 2628 CJ Delft, The Netherlands}
\author{Lars R. Schreiber}
\email{lars.schreiber@physik.rwth-aachen.de}
\affiliation{JARA-FIT Institute for Quantum Information, Forschungszentrum J\"ulich GmbH and RWTH Aachen University, Aachen, Germany}

\title{Singlet-triplet oscillations in multivalley Si double quantum dots} 

\begin{abstract}
Charge separation from the $(4,0)$ to the $(3,1)$ state in a Si/SiGe double quantum dot 
is commonly used for initialization of spin qubits and Pauli-spin-blockade readout. It was used in recent experiments involving creation of the $(3,1)$ singlet, and subsequent shuttling of one of the electrons.  We present a theoretical description of the process of charge separation and singlet-triplet mixing, arriving at expressions for the singlet return probability that take into account experimentally observed finite probabilities of the creation of singlets with various patterns of valley occupations. In our analysis we focus on magnetic fields for which the electron spin Zeeman splitting is close to the valley splitting in one of the dots, when the spin-valley coupling causes a strong renormalization of the frequency of oscillations of singlet return probability. The latter effect has been recently used to perform valley splitting mapping by shuttling of one quantum dot to various locations with respect to the other. We give a detailed description of singlet-triplet dynamics near these spin-valley resonances and compare the results of calculations with measurements on double quantum dots in two distinct Si/SiGe heterostructures. Comparison of theory with experiments in which the presence of a few valley occupation patterns is visible, gives insight into the valley dependence of $g$-factors in these structures, providing support for a recently proposed theoretical model of this dependence. We also discuss how dephasing of singlet return probability oscillations near the spin-valley resonances is affected by valley splitting fluctuations caused by electric field noise. 
\end{abstract}

\maketitle

\section{Introduction}
Electron spin qubits in silicon-based quantum dots - either Si/SiGe or SiMOS heterostructures - have achieved the level of gate errors and initialization/readout fidelities \cite{Yoneda_NN18,Struck_NPJQI20,Struck_SR21,Noiri_Nature22,Mills_SA22,Xue_Nature22,Philips_Nature22,Fuentes_arXiv25} compatible with the scalability of a quantum computer based on them. Electrons in silicon have weak spin-orbit coupling, making their spins less sensitive to charge noise and their $g$-factors less random than those of hole spins in silicon or germanium-based quantum dots \cite{Fang_MQT23}. Weak spin-photon coupling following from small spin-orbit interaction strength makes long distance coherent coupling between electron spin qubits challenging \cite{Samkharadze_Science18,Borjans_Nature20}. Architectures based on long-distance coherent shuttling of electrons in silicon were proposed  \cite{Langrock_PRXQ23,Boter_PRAPL22,Ginzel_PRB24,Kunne_NC24} to obviate the need for fine-tuning of spin-photon coupling, and rapid progress in long-distance shuttling of electrons \cite{Seidler_NPJQI22,Zwerver_PRXQ23,Xue_NC24} and coherent shuttling of electron spin qubits \cite{Yoneda_NC21,DeSmet_NN25,Struck_NC24,Foster_NPJQI25,Matsumoto_arXiv25,Krzywda_arXiv26} has been made recently. However, the use of electrons in Si as spin qubits comes with another drawback: the presence of two low-energy valley states \cite{Friesen_PRB07,Saraiva_PRB09,Friesen_PRB10,Zwanenburg_RMP13} having energy splitting that is typically $< \! 100$ $\mu$eV in Si/SiGe structures. The mere presence of two valley states (not to mention the possibility of electron in a quantum dot occupying both of them with finite probability) poses challenges for qubit initialization, control, and readout \cite{Kawakami_NN14,Philips_Nature22}, motivating intense research aimed at understanding the physics controlling the magnitude of valley splitting \cite{Wuetz_NC22,Lima_MQT23,Losert_PRB23,Klos_AS24,Thayil_arXiv24,Woods_NPJQI24}. Of particular significance for this Paper is the existence of spin-valley coupling enabled by the interplay of spin-orbit coupling and valley-orbit mixing by interface roughness \cite{Yang_NC13,Corna_NPJQI18,Zhang_PRL20,Huang_NPJQI21}. This coupling strongly mixes spin and valley degrees of freedom whenever the Zeeman splitting for an electron in a given quantum dot, $\EZ{}$, is close to the valley splitting, $\EVS$, in this dot, i.e.~when the qubit is close to the spin-valley resonance.

The possibility of exciting a silicon spin qubit to a higher-energy valley state during electron motion has to be carefully optimized against when shuttling qubits \cite{Langrock_PRXQ23,Losert_PRXQ24,David_arXiv24,Nemeth_arXiv24}. On the other hand, shuttling of spin superposition states has recently been shown to be a useful tool for characterization of spatially random valley splitting over large areas of Si/SiGe quantum well \cite{Volmer_NPJQI24,Volmer_arXiv26}. In the experiments \cite{Struck_NC24,Volmer_NPJQI24} a spin singlet is separated spatially in a double quantum dot (DQD), and then one of the dots is moved to a desired location, at distance $d$ from its initial position, using conveyor-mode charge shuttling \cite{Langrock_PRXQ23,Struck_NC24}. The dynamics in the singlet ($S$) and unpolarized triplet ($T_0$) subspace are sensitive to difference of spin splittings in the two dots, given by $\Delta g \mu_B B$, where $\Delta g \! \lesssim \! 10^{-3}$ is the typical magnitude of $g$-factor difference for two Si/SiGe quantum dots \cite{Kawakami_NN14,Ferdous_NPJQI18,Volmer_NPJQI24}. 
When the magnetic field is such that the spin qubit in the shuttled dot is close to the spin-valley resonance, spin-valley coupling leads to a renormalization of the Zeeman splitting of the qubit \cite{Jock_NC22,Struck_NC24,Jacobson_arXiv26}. 
While such a renormalization amounts to a small ($<10^{-3}$) fraction of $\EZ{}$ of a single spin, it leads to a $\sim 100 \%$ renormalization of singlet-unpolarized triplet splitting given by $\Delta g \mu_B B$. Subsequent shuttling of one of the dots back to its original position, followed by performing the Pauli spin blockade (PSB) readout of singlet/triplet state of two spins in the DQD, allows a measurement of the singlet-triplet precession that is very sensitive to the value of $|\EVS-E_{Z}|$ at the position $d$, at which most of the spin dynamics took place. This sensitivity leads to the presence of characteristic discontinuities in $(E_Z,d)$ map of the singlet return probability, and thus to reconstruction of $\EVS(d)$ dependence \cite{Volmer_NPJQI24,Volmer_arXiv26}.

In this paper we give a detailed theoretical analysis of the singlet-return-probability signal in a setup used in \cite{Struck_NC24,Volmer_NPJQI24,Volmer_arXiv25}, in which a charge separation from $(4,0)$ to $(3,1)$ state is used to initialize a spatially delocalized spin singlet. Our main focus is on ranges of magnetic fields for which the spin splitting in one of the dots is close to the valley splitting, so the spin-valley couplings in both dots are the crucial interactions in our DQD model \cite{Hao_NC14,Hwang_PRB17,Jock_NC22,Liu_PRAPL21,Cai_NP23,Jacobson_arXiv26}.
We give particular attention to the possibility of creation of finite occupation of more than one valley occupation pattern for $(3,1)$ singlet that leads to the presence of additional frequencies in the measured $P_S(t)$ signal. The presence of such additional frequencies was noted in the first experiments \cite{Struck_NC24}, but subsequent analysis of spin-valley resonances in \cite{Volmer_NPJQI24} was done for one chosen frequency branch, while the results for $\EVS(d)$ mapping were independent on the choice of the branch, as we explain in detailed manner in this paper. Comparison of theoretical predictions with detailed measurements of $B$ dependence of $P_S(t)$ signals (actually, the frequency dependence of Fourier-transformed singlet return probability signals) near the spin-valley resonances in devices from two distinct Si/SiGe heterostructures, used in \cite{Volmer_NPJQI24} and \cite{Volmer_arXiv26}, respectively, gives support to the recently proposed model of valley-dependence of $g$-factors in Si/SiGe quantum dots \cite{Woods_arXiv24,Woods_arXiv25}. Furthermore, we discuss the dephasing of the $P_S(t)$ signal near the spin-valley resonances caused by fluctuations in both $\EZ{}$ and $\EVS$, and uncover that in the vicinity of the spin-valley resonance it is the valley splitting noise that can dominate the decay of the signal, especially in isotopically purified samples. 

The paper is structured in the following way. In Section \ref{sec:model} we describe the model for the DQD under consideration, focusing in Sec.~\ref{sec:Hsingle_el} on single-electron Hamiltonian in a silicon DQD, and proceeding in Sec.~\ref{sec:H_singlets} to description of the model describing the physics of 4 electrons in $(4,0)-(3,1)$ charge occupations of the DQD. We provide a detailed description of the spin-valley mixing for all the possible valley occupations of $(3,1)$ singlets that can be created by detuning sweep leading to transfer of one electron from the left dot to the right one. In Sec.~\ref{sec:nonadiabatic} we focus on the possibility of charge separation leading to initilization of mixture of two or more $(3,1)$ singlets with distinct valley occupation patterns. We first briefly discuss \ref{sec:nonadiabatic_mechanisms} in the possible reasons for non-perfectly adiabatic character of charge separation during detuning sweep from $(4,0)$ to $(3,1)$, and then discuss in Sec.~\ref{sec:PS_decomposition} the impact of such non-adiabaticity on the singlet return probability signal observed in the PSB measurement. Since the final result of this Section is that the signal is expected to be a sum of signals corresponding to the singlets that are created with nonzero amplitude, in Section \ref{sec:ST_oscillations} we analyze the magnetic field dependence of the signal for a given singlet, paying particular attention to the behavior near one of spin-valley resonances. 
In Section \ref{sec:dephasing} we include in these calculations the effects of fluctuations of spin and valley splittings in the two dots. 
With a theoretical model of the singlet return probability in the presence of occupation of multiple valley states and dephasing, we proceed in Section \ref{sec:experiment_comparison} to the analysis of experimental results obtained in DQDs in two distinct Si/SiGe structures used in recent experiments on long-distance shuttling of one of the dots \cite{Volmer_NPJQI24,Volmer_arXiv25}. In Sec.~\ref{sec:exp} we describe the experimental results and the parameters obtained from fitting the theoretical model to them, and in Sections \ref{sec:dephasing_exp} and \ref{sec:g_factor} we discuss, respectively, the implications of these observations for magnitude of valley splitting fluctuations, and valley-dependence of $g$-factors in Si/SiGe quantum dots.

\section{The Model} \label{sec:model}
\subsection{Single electron in a double quantum dot} \label{sec:Hsingle_el}
We describe a single electron in a silicon DQD taking into account the lowest energy orbital states in the left (L) and right (R) dot, the two lowest energy valley eigenstates per dot, labeled by $\nu_{D}$ with $\nu \!= \!g$ $(e)$ for the ground (excited) valley state in dot $D=L/R$, and the electron's spin degree of freedom $\sigma \! = \! \uparrow,\downarrow$. The single-electron basis is thus spanned by $\ket{\nu_D \sigma}$ states. The valley eigenstates in a given dot, $\ket{\nu_D}$, are eigenstates of the valley coupling Hamiltonian \cite{Friesen_PRB07,Friesen_PRB10,Saraiva_PRB09,Saraiva_PRB11,Culcer_PRB10,Zwanenburg_RMP13,Thayil_PRB25}:
\beq
\hat{H}_v = \left( \begin{array}{cc}  
0 & \Delta_D   \\
\Delta^*_D &  0 
\end{array}
\right) \,\, ,
\eeq
written in the basis of $\{\ket{+z},\ket{-z}\}$ states (Bloch waves at minima of conduction band along the $[001]$ crystal direction located at $\pm k_z \! \approx \! 0.85 2\pi/a_0$ where $a_0$ is the lattice constant of Si), in which  $\Delta_D \! \equiv \! |\Delta_D|e^{-i\phi_D}$ is the valley coupling, the values of which are determined by the structure of the interface (Si-SiGe interface in case of Si/SiGe structures) on an atomic lengthscale \cite{Friesen_PRB07,Friesen_PRB10,Saraiva_PRB09,Saraiva_PRB11,Culcer_PRB10,Zwanenburg_RMP13,Thayil_PRB25}. The splitting between the valley eigenstates in dot $D$ is thus given by
$\EVD{D} = 2|\Delta_D|$.
According to the current theoretical \cite{Wuetz_NC22,Lima_MQT23,Losert_PRB23,Thayil_PRB25} understanding that is supported by the newest experiments on statistic of valley splitting \cite{Volmer_NPJQI24,Volmer_arXiv26,Marcks_NC25}, the value of $\Delta_D$ in Si/SiGe quantum dots is determined by alloy disorder. An important result of this model is the observation that for two dots (labeled $L$ and $R$)  for which the ground state envelope functions have little overlap, the real and imaginary parts of $\Delta_L$ and $\Delta_R$ are Gaussian-distributed and approximately uncorrelated random variables \cite{Losert_PRB23}.  

The Hamiltonian can be written as a sum of spin-diagonal terms $\hat{H}_{0}$, the Zeeman splitting term $\hat{H}_Z$, and the spin-valley coupling $\hat{H}_{sv}$:
\beq
\hat{H}_{\mathrm{se}} = \hat{H}_0 + \hat{H}_Z+ \hat{H}_{sv} \,\, .  \label{eq:Hse}
\eeq
The spin-diagonal term is given by
\begin{align}
\hat{H}_{0} & = \epsilon_{L}\sum_{\nu} \ket{\nu_L}\bra{\nu_L} + \epsilon_{R}\sum_{\nu}\ket{\nu_R}\bra{\nu_R} +\nonumber\\
& \EVD{L}\ket{e_L}\bra{e_L} +  \EVD{R}\ket{e_R}\bra{e_R} + \nonumber\\
& 
\sum_{\nu_R,\mu_L}\left( t_{\mu_L \nu_R} \ket{\mu_L}\bra{\nu_R} + \mathrm{h.c.} \right) \,\, , \label{eq:H0}
\end{align}
where $\epsilon_D$ are the ground state orbital energies in each QD, $\EVD{D}$ are valley splitting energies in dot $D$, and $t_{\mu_L \nu_R}$ are tunnel couplings between  all the valley eigenstates in the two dots. For less cluttered notation we will omit from now on the $L$ and $R$ subscripts, and use notation $t_{\mu\nu}$ for tunnel couplings, with the first valley subscript referring to the $L$ dot, and the second referring to the $R$ dot. 
Since the ``bare'' tunneling $t_0$ is diagonal in the $\ket{\pm z}$ valley states \cite{Culcer_PRB10}, the tunnel couplings between valley eigenstates in the two dots are determined by the overlaps of the respective eigenstates in the two dots \cite{Tariq_NPJQI22,Zhao_PRAPL22}:
\begin{align}
t_{ee} = t_{gg} & = \frac{1}{2}t_0( 1 + e^{-i\Delta\phi} ) \,\, , \nonumber\\
t_{eg} = t_{ge} & = \frac{1}{2}t_0( 1 - e^{-i\Delta\phi} ) \,\, , \label{eq:tmunu}
\end{align}
where $\Delta\phi \! = \! \phi_L - \phi_R$ is the valley phase difference between the two dots.

The valley- and dot-dependent spin splittings are described by
\beq
\hat{H}_Z =  \frac{1}{2} \sum_{D,\nu} \EZ{\nu_D} \ket{\nu_D}\bra{\nu_D}  \hat\sigma_{z}^D  \,\, , \label{eq:Hs}
\eeq
in which $\EZ{\nu_D} = g_{\nu_D}\mu_B B_{z}^D$ are dot- and valley-specific magnetic Zeeman splittings, with $B_{z}^{D}$ accounting for the possible presence of a magnetic field gradient making the $B$ fields at position of each dot distinct, and $g_{\nu_D}$ being the dot- and valley-dependent $g$-factor \cite{Veldhorst_PRB15,Ruskov_PRB18,Woods_arXiv24,Woods_arXiv25}. When the influence of hyperfine coupling with the nuclei of $^{29}$Si (or nuclei of $^{73}$Ge with which the electron wavefunction has finite overlap \cite{Struck_NPJQI20,Cvitkovich_PRAPL24}) needs to be considered, dot-specific Overhauser fields, $h_z^D$, should be added to the spin splittings $\EZ{\nu_D}$. 

Finally, the spin-valley coupling is described by
\beq
\hat{H}_{sv} = \sum_{D=L,R} \left( v_D \ket{g_D \uparrow}\bra{e_D \downarrow} + v'_D\ket{g_D \downarrow}\bra{e_D \uparrow}  + \mathrm{h.c.} \right) \,\, , \label{eq:Hsv}
\eeq
in which $v_{i}$ and $v'_{i}$ are the spin-valley couplings \cite{Yang_NC13,Huang_PRB14}. Their previously reported values range from $20$ neV \cite{Yang_NC13} to $100-200$ neV \cite{Hao_NC14,Hwang_PRB17,Jock_NC22,Jacobson_arXiv26} in SiMOS devices, and between $\approx 20-50$ neV \cite{Cai_NP23,Jacobson_arXiv26} and $60-80$ neV \cite{Volmer_NPJQI24} in Si/SiGe. 
In this paper we will not only revisit the measurement from  \cite{Volmer_NPJQI24}, but also report $v$ in the $100-200$ neV range measured in a Si/SiGe heterostructure distinct from the one investigated in that reference. Note that since $\EZ{}$ and $\EVS{}$ are defined to be positive quantities here, the spin-valley term $v'$ couples the states that are never close in energy, so that the presence of this term can be disregarded. 

\optional{Note that we have neglected spin-flip tunneling in the above Hamiltonian. The maximum  values of spin-flip tunnel coupling, $t_{\mathrm{sf}} \sim 0.1$ $\mu$eV, were reported in SiMOS devices \cite{Harvey-Collard_PRL19}, and can be considered an upper bound for values in Si/SiGe structures. Below we will explain why these can be neglected - Because the $S-T_{-}$ anticrossing is surely passed diabatically?}

\subsection{Hamiltonian for a system initialized by charge separation starting from the $(4,0)$ charge configuration} \label{sec:H_singlets}
We consider now a DQD initialized in the ground state with $4$ electrons loaded into the L dot, i.e.~in the $(4,0)$ charge configuration that is stable for detuning $\epsilon \! \equiv \! \epsilon_L - \epsilon_R \! \ll \! -|t_0|$. In this state, both valleys are occupied by two electrons, and we can think of each valley as being occupied by a spin singlet \cite{Cai_NP23}, and thus we label this state as $\ket{S(4,0)}$, or $\ket{S_0}$ for shorter notation.

\begin{figure}[tb]
\includegraphics[width=\columnwidth]{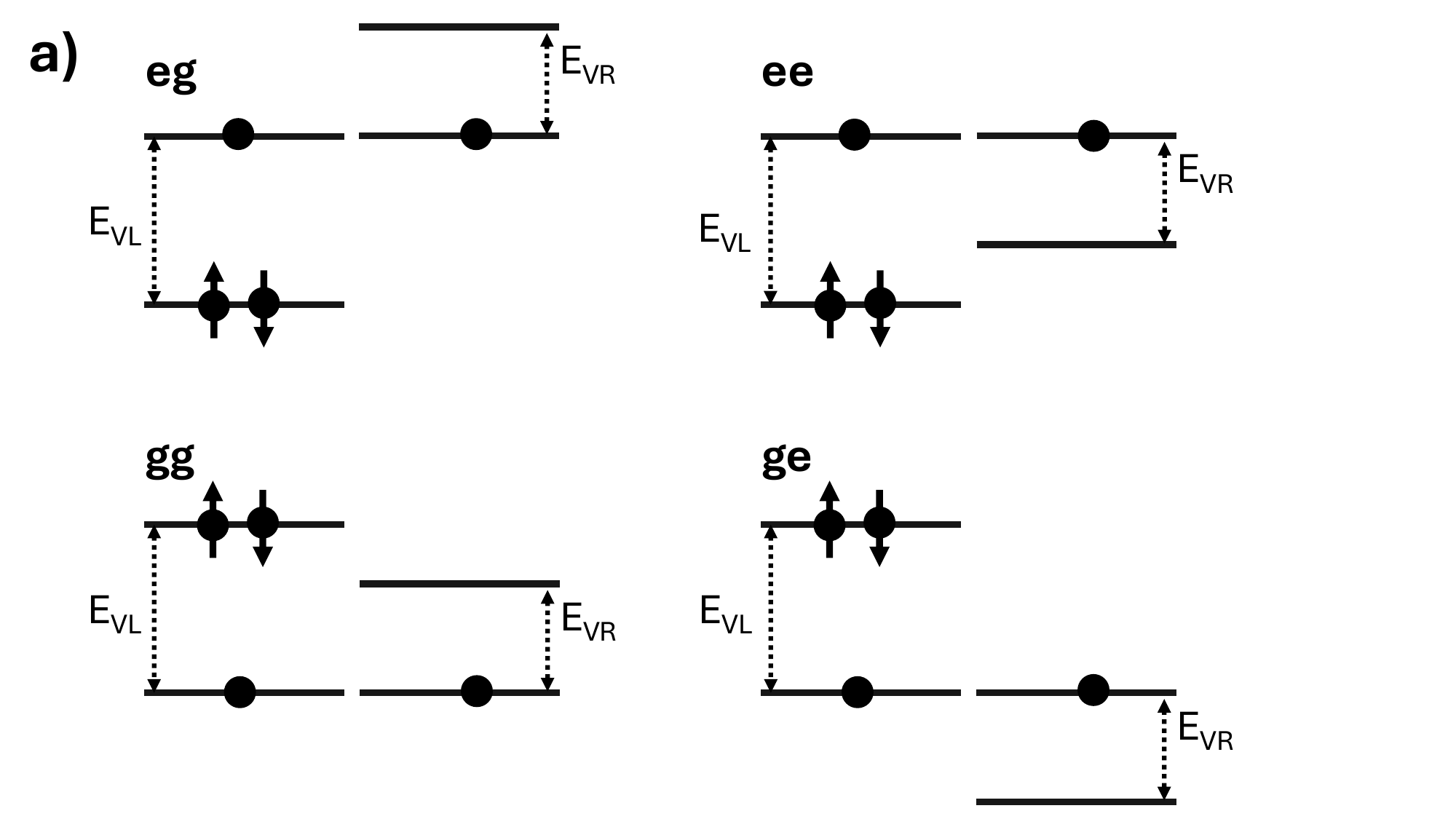}

\vspace{1.0cm}

\includegraphics[width=\columnwidth]{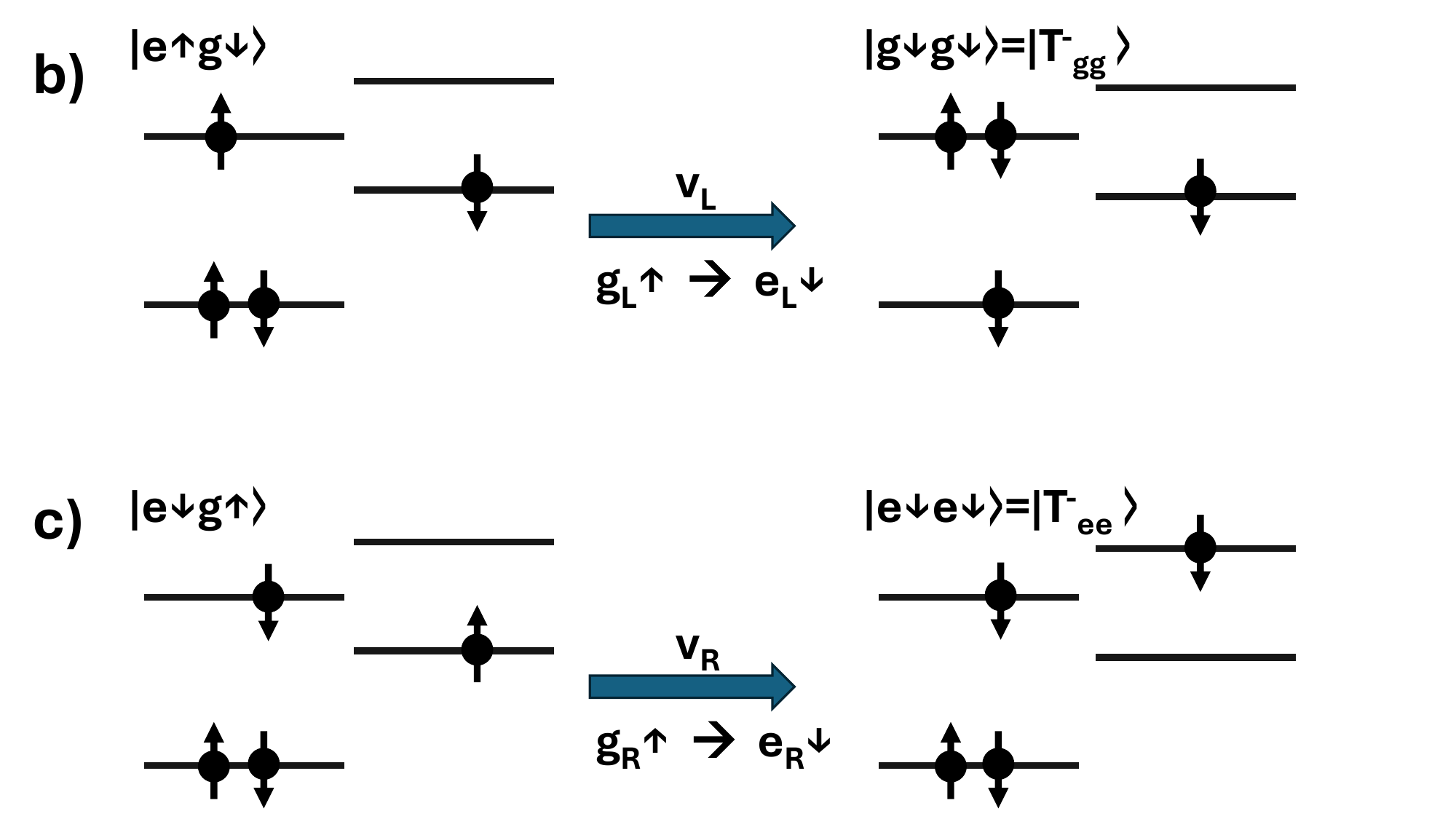}
    \caption{(a) Occupations of valley states in $L$ and $R$ dots in $(3,1)$ charge configuration. The states in which the $\mu$ valley state in dot $L$ and $\nu$ valley state in dot $R$ are  occupied by a single electron are labeled $\mu\nu$. 
    For $\EVD{L}>\EVD{R}$, when going clockwise from the $eg$ state we encounter the states that can be created by interdot tunneling as the detuning is increased. For a detuning sweep that is adiabatic with respect to the $t_{eg}$ tunnel coupling only the $S_{eg}$ state is created; if the sweep is not adiabatic, the subsequent state, $S_{ee}$, can be created due to $t_{ee}$ tunnel coupling, etc. (b,c) Illustration of states with ``eg'' valley occupation that are spin-valley coupled. The ``eg'' singlet with $(3,1)$ charge occupation is written as $\ket{S_{eg}} \! =\! (\ket{e\uparrow g\downarrow} - \ket{e\downarrow g\uparrow})/\sqrt{2}$, and in (b) we see that spin-valley coupling in the left dot, $v_L$, mixes $\ket{e\uparrow g\downarrow}$ with the $\ket{T^{-}_{gg}}$ polarized triplet, while in (c) we see how the spin-valley coupling in the right dot, $v_R$, mixes $\ket{e\downarrow g\uparrow}$ with the $\ket{T^{-}_{ee}}$ polarized triplet.
     }
		\label{fig:labelling}
\end{figure}

\begin{figure}[tb]
\includegraphics[width=\columnwidth]{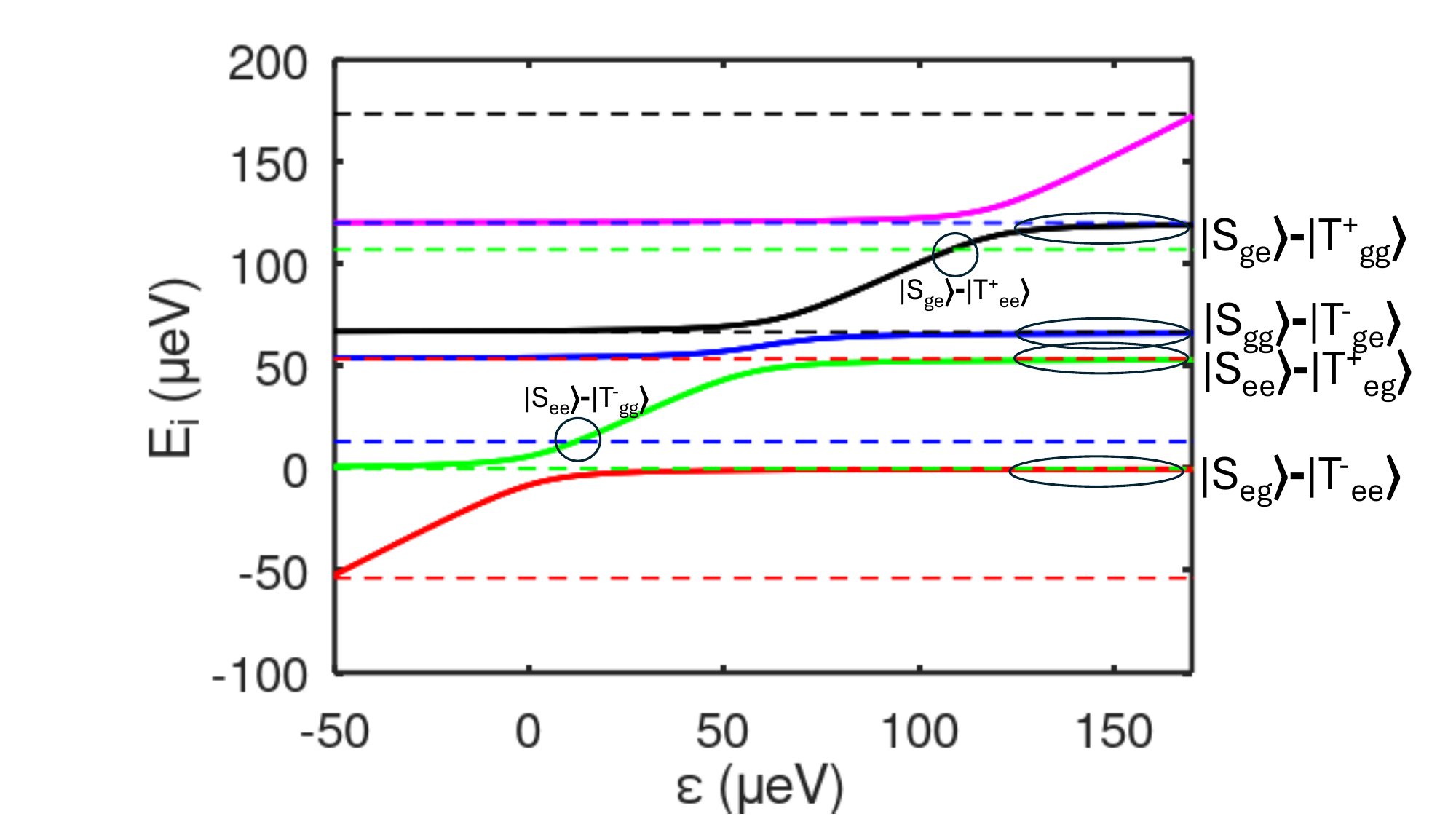}
    \caption{Energies of singlet states (solid lines) and  energies of spin-polarized $(3,1)$ triplet states (dashed lines) that can be mixed by spin-valley coupling with the singlets, calculated for $2\mu_B B \! =\! \EVD{R}$. Some of the spin-valley anticrossings (funnels) that occur at detunings at which the singlet is a superposition of $(4,0)$ and $(3,1)$ charge states are marked with circles. The 4 regions in which a $(3,1)$ singlet is mixed with a polarized triplet at large detuning are marked by ellipses. These are the spin-valley resonances at which the spin-valley coupling in the right dot affects the singlet dynamics in the far detuned regime, see the lines corresponding to odd $r$ in Table \ref{tab:resonances}. Another set of 4 such regions, corresponding to even $r$ in this Table, appears when $2\mu_B B \! =\! \EVD{L}$.
    }
		\label{fig:all_funnels}
\end{figure}

This state is tunnel-coupled via $t_{\mu \nu}$ to four states in $(3,1)$ charge configuration. For each of these states, one of the valleys in the left (L) dot, $\mu$ is singly occupied, and the joint state of the electron in this valley and the electron in $\nu_R$ valley in the right (R) dot is a singlet. We will thus label these states as $\ket{S_{\mu \nu}}$, specifying the singly-occupied valley states in each of the dots, see Fig.~~\ref{fig:labelling}a for illustration. All the singlets can also be labelled with $k=0\ldots 4$ index, with $k=1,2,3,4$ corresponding to $\mu\nu=eg,ee,gg,ge$, respectively.
We thus have the following Hamiltonian in the basis of $\{\ket{S_0},\ket{S_{eg}},\ket{S_{ee}},\ket{S_{gg}},\ket{S_{ge}}\} \! \equiv \! \{\ket{S_0},\ket{S_1},\ldots,\ket{S_4}\}$ given by
\beq
\hat{H} = \left( \begin{array}{ccccc}
\epsilon        &  \sqrt{2}t_{eg}    & \sqrt{2}t_{ee}   & \sqrt{2}t_{gg} & \sqrt{2}t_{ge}   \\
\sqrt{2}t^{*}_{eg}  &   0   & 0         &  0        &  0   \\
\sqrt{2}t^{*}_{ee}  &   0   & \EVD{R}    &  0        &  0   \\
\sqrt{2}t^{*}_{gg}  &   0   & 0         &  \EVD{L}   &  0   \\
\sqrt{2}t^{*}_{ge}  &   0   & 0         &  0        &  \EVD{L}+\EVD{R}   
\end{array} \right)  \,\, , \label{eq:Hsinglets}
\eeq
where $\epsilon \! =\! \epsilon_L-\epsilon_R + \EVD{L} + U_L$ (with $U_L$ being the Coulomb energy difference between $(4,0)$ and $(3,1)$ charge configurations), and an energy shift of $3\epsilon_L+\epsilon_R+\EVD{L}$ was applied in order to make the  anticrossing  of the $\ket{S_0}$ and $\ket{S_{eg}}$ states occur at $\epsilon\! =\! 0$. 
The eigenenergies of the above Hamiltonian as functions of $\epsilon$ are plotted as solid lines in Fig.~\ref{fig:all_funnels} for $\EVD{R} \! = \! 53.5$ $\mu$eV and $\EVD{L} \! =\! 66.6$ $\mu$eV, measured for one of  DQDs investigated in \cite{Volmer_NPJQI24}, with all the tunnel couplings $t_{\mu\nu}$ set to a common value of $t_c \! =\! 10$ $\mu$eV typical for Si/SiGe DQDs \cite{Cai_NP23}.

The Hilbert space of states with $(3,1)$ charge configuration built out of single-electron states used in Hamiltonian \eqref{eq:Hse} is 16-dimensional. Apart from the four $\ket{S_{\mu\nu}}$ states appearing in the above Hamiltonian, it contains 4 groups of $\ket{T_{\mu\nu}^{p}}$ triplets with $p\! = \! 0, \pm 1$ denoting the spin projection on the $z$ axis, having energies $E(T_{\mu\nu}^p) \! =\! E(S_{\mu\nu}) + p \AEZ{\mu\nu}$, where 
\beq
\AEZ{\mu\nu} = \frac{1}{2}(\EZ{\mu_L} +\EZ{\nu_R}) \,\, . \label{eq:AEZ}
\eeq 
The interdot differences in spin splittings, defined as
\beq
\DEZ{\mu\nu}  = \EZ{\mu} - \EZ{\nu} \,\, ,  \label{eq:DEZ}
\eeq
couple $\ket{S_{\mu\nu}}$ states to $\ket{T^{0}_{\mu\nu}}$ states \cite{Taylor_PRB07}, while  the spin-valley terms $v_{D}$, $v'_{D}$ are coupling $\ket{S_{\mu\nu}}$ and $\ket{T^0_{\mu\nu}}$ states with $\ket{T_{\mu\bar{\nu}}^{\pm}}$ and $\ket{T_{\bar{\mu}\nu}^{\pm}}$ states with $p\!=\!\pm 1$ (see Fig.~\ref{fig:labelling}(b,c) for illustration in the case of the $\ket{S_{eg}}$ state), where we have introduced the notation  in which $\bar{\mu}$ denotes $e(g)$ if $\mu \! =\! g(e)$.

These spin-valley interactions mixing $(3,1)$ singlets with the {\it polarized} triplets are relevant only at values of $\epsilon$ at which the unperturbed energies of the states differ by a value that is $\lesssim \! v_D$. These are the spin funnels analyzed in \cite{Liu_PRAPL21,Cai_NP23} in the regime of sizable singlet-triplet exchange splitting, $J_{\mu\nu}$, i.e.~for detuning with respect to a given anticrossing being of the order of the respective tunnel coupling. An example of such funnels at $\ket{S_{ee}}-\ket{T_{gg}^{-}}$ and  $\ket{S_{ge}}-\ket{T_{ee}^{+}}$ anticrossings are marked by circles in Fig.~\ref{fig:all_funnels}.
On the other hand, mixing of $\ket{S_{\mu\nu}}$ with $\ket{T_{\mu\nu}^0}$ via $\DEZ{\mu\nu}$ becomes relevant when $\epsilon$ is so large that we are in  $J_{\mu\nu} \! \ll \! \DEZ{\nu\mu}$ regime in which $S-T_0$ precession due to Zeeman splitting difference can occur. 
 \optional{From the Hamiltonian \eqref{eq:Hsinglets} we have $J_{\nu\mu} \approx 2 t_{\nu\mu}^2/(\epsilon-\epsilon_{\nu\mu})$ in which $\epsilon_{\nu\mu} \! =\! 0$, $E_{R}^{V}$, $E_{V}^{L}$, $E_{R}^{V}+E_{V}^{L}$, for $\nu\mu \! =\! eg$, $ee$, $gg$, and $ge$, respectively. For $t_{\nu\mu} \! \sim\! 10$ $\mu$eV and $\DEZ{}\lesssim 0.1$ $\mu$eV this leads to condition $(\epsilon-\epsilon_{\nu\mu}) \! \gtrsim 1$ meV. Since the above formula for $J$ is well known to overestimate $J$ in the far-detuned regime, in which $J\propto e^{-\epsilon/\epsilon_0}$ is typically measured \cite{Dial_PRL13,Cai_NP23}, a more precise estimate of $\epsilon$ above which $J$ becomes smaller than $\DEZ{}$ is best left to be determined experimentally.  }
If the Zeeman splitting of the polarized triplets with respect to the $S/T_0$ states is close to valley spllitting in any of the dots, at large $\epsilon$ each $\ket{S_{\nu\mu}}$ state is mixed with both the $\ket{T_{\nu\mu}^0}$ state, and the respective polarized triplet corresponding to a distinct valley occupation pattern. 
In Fig.~\ref{fig:all_funnels} we schematically mark such regions by ellipses.  Our focus here is on the dynamics of states coupled by Zeeman splitting difference $\DEZ{}$, {\it and also the spin-valley coupling $v_D$} in this far-detuned regime. 

In Table \ref{tab:resonances} we list the polarized triplets that are appreciably mixed with the respective singlets at magnetic fields for which the Zeeman energy $\AEZ{\nu\mu} \! \approx \! 2\mu_B B$ is close to either $\EVD{L}$ or $\EVD{R}$ valley splitting. Let us give now the Hamiltonians governing the dynamics in subspaces containing each of the $\ket{S_{\nu\mu}}$ singlets. For further discussion it is convenient to use the basis of states $\ket{\nu\uparrow\mu\downarrow} \! \equiv \! (\ket{T^0_{\nu\mu}}+\ket{S_{\nu\mu}})/\sqrt{2}$ and $\ket{\nu\downarrow\mu\uparrow} \! \equiv \! (\ket{T^0_{\nu\mu}}-\ket{S_{\nu\mu}})/\sqrt{2}$.   
Using such a 
$\{\ket{e\uparrow g\downarrow},\ket{e\downarrow g\uparrow},\ket{T^{-}_{ee}},\ket{T^{-}_{gg}}$ basis we have the Hamiltonian describing the coupling of the $\ket{S_{eg}}$ singlet to the respective triplets given by 
\beq
\hat{H}_{eg} = \left( \begin{array}{cccc}
\frac{\DEZ{eg}}{2} & 0  & 0 & v_L \\
0  &  -\frac{\DEZ{eg}}{2}   & v_R & 0 \\
0 & v_R^{*} & \EVD{R}-\AEZ{ee} & 0 \\
v_L^{*} & 0 & 0 & \EVD{L}-\AEZ{gg}
\end{array} 
\right )\,\, . \label{eq:Heg}
\eeq
Analogously, for $\ket{S_{ge}}$ we use the basis of $\{\ket{g\uparrow e\downarrow},\ket{g\downarrow e\uparrow},\ket{T^{+}_{gg}},\ket{T^{+}_{ee}}\}$ to obtain
\begin{align}
\hat{H}_{ge}&  = \epsilon_{ge} \mathds{1}+
\left( \begin{array}{cccc}
\frac{\DEZ{ge}}{2}  & 0  & v_R & 0 \\
0  &  -\frac{\DEZ{ge}}{2}   & 0 & v_L \\
v_R^{*} & 0 & \AEZ{gg}-\EVD{R} & 0 \\
0 & v_L^{*} & 0 & \AEZ{ee}-\EVD{L} 
\end{array} 
\right ) \,\, . \label{eq:Hge}
\end{align}
in which $\epsilon_{ge} \!= \! \EVD{L}+\EVD{R}$. 

Note that for the system initialized in $\ket{S_{eg}}$ or $\ket{S_{ge}}$, the dynamics due to the above Hamiltonians occur in disjoint four-dimensional subspaces. On the other hand, for the system initialized in $\ket{S_{ee}}$ or $\ket{S_{gg}}$, the dynamics occurs in the same six-dimensional subspace spanned by $\{\ket{e\uparrow e\downarrow},\ket{e\downarrow e\uparrow},\ket{g\uparrow g\downarrow},\ket{g\downarrow g\uparrow},\ket{T^{+}_{eg}},\ket{T^{-}_{ge}}\}$:
\begin{widetext}
\beq
\hat{H}_{ee/gg} = \left( \begin{array}{cccccc}
\EVD{R}+\frac{\DEZ{ee}}{2} & 0  & 0 &  0 &  v_R & v_L \\
0  &  \EVD{R}-\frac{\DEZ{ee}}{2}  & 0 & 0 & 0 & 0 \\
0 & 0 & \EVD{L}+\frac{\DEZ{gg}}{2} & 0 & 0 & 0 \\
0 & 0 & 0 & \EVD{L}-\frac{\DEZ{gg}}{2} & v_L & v_R \\ 
v_R^* & 0 & 0 & v_L^* & \AEZ{eg} & 0 \\
v_L^* & 0 & 0 & v_R^* & 0 & \EVD{L}+\EVD{R}-\AEZ{ge} 
\end{array} 
\right )\,\, . \label{eq:Heegg}
\eeq
\end{widetext}
However, unless the valley splittings in the two QDs are very close to each other, with $|\EVD{L}-\EVD{R}| \! \lesssim |v_D| \! \lesssim \! 0.1$ $\mu$eV, for any given value of magnetic field the spin-valley interaction in only one of the dots leads to significant mixing of states. Specifically, the $\ket{e\uparrow e\downarrow}$ state mixes with $\ket{T^{+}_{eg}}$ ($\ket{T^{-}_{ge}}$) when   $\EZ{eg} \! \approx \! \EVD{R}$ ( $\EZ{ge} \! \approx \! \EVD{L}$), while the $\ket{g\downarrow g\uparrow}$ state mixes with $\ket{T^{-}_{ge}}$ ($\ket{T^{+}_{eg}}$) when   $\EZ{ge} \! \approx \! \EVD{R}$ ( $\EZ{eg} \! \approx \! \EVD{L}$). When one of these states is significantly mixed with the respective triplet, the energy of the other is only weakly renormalized by spin-valley coupling with a triplet that is separated by an energy $\gg \! v_D$.
The same holds for $\hat{H}_{eg}$ and $\hat{H}_{ge}$ Hamiltonians from Eqs.~(\ref{eq:Heg}) and (\ref{eq:Hge}): if $|\EVD{L}-\EVD{R}| \! \gg \! |v_D|$, which is a generic situation for a DQD in Si/SiGe structure, only one of $\ket{\nu\uparrow \mu\downarrow}$ and $\ket{\nu\downarrow \mu\uparrow}$ states is appreciably mixed with the respective polarized triplet, while the shift of the energy of the other one can be treated in second order perturbation theory.

Consequently, in order to understand the dynamics of the system initialized in any of the $\ket{S_{\mu\nu}}$ states in the far-detuned regime, in the generic situation in which $|\EVD{L}-\EVD{R}| \! \gg \! |v_D|$ we only need to consider dynamics in one of four disjoint three-dimensional subspaces. Depending on the singlet of interest $\ket{S_{\nu\mu}}$, and whether we are interested in magnetic field range in which $2\mu_B B \! \approx \! \EVD{R}$ of $\EVD{R}$, we choose one of the $8$ case listed in Table \ref{tab:resonances}, and in the basis of $\{\ket{\mu\uparrow \nu\downarrow},\ket{\mu\downarrow \nu\uparrow},|T^p_{\alpha\beta}\rangle\}$ we have 
\beq
\hat{H}_{\mu\nu,\alpha\beta}  = \left( \begin{array}{ccc}
\DEZ{\nu\nu}/2  & 0 & v_D(1+\xi)/2 \\
0  &  -\DEZ{\mu\nu}/2   & v_D(1-\xi)/2   \\
v^*_D(1+\xi)/2  & v^*_D(1-\xi)/2 & p(\AEZ{\alpha\beta}-\EVD{D})
\end{array} 
\right ) \,\, . \label{eq:Hk_ud}
\eeq
where $\xi \! =\! \pm 1$, $p\! = \! \pm 1$, and $D \! =\! L$ or $R$ are taken from the row of Table \ref{tab:resonances} corresponding to the respective set of $\mu\nu$ and $\alpha\beta$ valley indices (equivalently, respective index $r \! =\! 1 \ldots 8$). Note that $\xi\! =\! 1$ ($-1$) means that it is the $\ket{\mu\uparrow \nu\downarrow}$ ($\ket{\mu\downarrow \nu\uparrow}$) that is coupled by $v_D$ to the $|T^p_{\alpha\beta}\rangle$ triplet, while $p \! =\! 1$  ($-1$) means that as $B$ field is increased, thus increasing $\AEZ{\alpha\beta}$, the energy of the polarized triplet is approaching the energy of the state in $\ket{S_{\mu\nu}}-\ket{T^{0}_{\mu\nu}}$ manifold coupled with the triplet from below (above). 

\begin{table}[tb]
 \centering
\begin{tabular}{|c|l|c|c|c|c|c|}
\hline	
$r$  &$S_{\mu\nu}$    & Resonance   & $T^p_{\alpha\beta}$ & $D$ & $\xi$ & $p$ \\ \hline
1&$S_{eg}$         & $\AEZ{ee}\approx \EVD{R}$  & $T_{ee}^{-}$  & $R$  & $-1$ & $-1$   \\ \hline
2&$S_{eg}$         & $\AEZ{gg}\approx \EVD{L}$  & $T_{gg}^{-}$  & $L$  & $+1$  & $-1$  \\ \hline
3&$S_{ee}$         & $\AEZ{ge}\approx \EVD{R}$  & $T_{eg}^{+}$  & $R$  & $+1$ & $+1$   \\ \hline
4&$S_{ee}$         & $\AEZ{ge}\approx \EVD{L}$  & $T_{ge}^{-}$  & $L$  & $+1$  & $-1$  \\ \hline
5&$S_{gg}$         & $\AEZ{ge}\approx \EVD{R}$  & $T_{ge}^{-}$  & $R$  & $-1$  & $-1$  \\ \hline
6&$S_{gg}$         & $\AEZ{eg}\approx \EVD{L}$  & $T_{eg}^{+}$  & $L$  & $-1$  & $+1$  \\ \hline
7&$S_{ge}$         & $\AEZ{gg}\approx \EVD{R}$  & $T_{gg}^{+}$  & $R$  & $+1$  & $+1$  \\ \hline
8&$S_{ge}$         & $\AEZ{ee}\approx \EVD{L}$  & $T_{ee}^{+}$  & $L$  & $-1$  & $+1$  \\ \hline
\end{tabular}
\caption{Parameters of the singlet-triplet mixing Hamiltonian from Eq.~\eqref{eq:Hk_ud} for all combinations of $\ket{S^{\nu\mu}}$ singlets and polarized $\ket{T_{\alpha\beta}^p}$ states for which spin-valley mixing resonance occurs at values of Zeeman splitting given in the third column. 
}
\label{tab:resonances}
\end{table} 

Depending on the parameters of the system, in order to precisely capture the frequency of singlet-triplet oscillations, sometimes we have to include the perturbative correction ($\sim |v_{D}|^2/(\EVD{L}-\EVD{R})$, where $D' \! =\ L(R)$ when $D\! =\! R(L)$) to the unperturbed energy of the state that is not mixed with its corresponding triplet in this Hamiltonian. In the following, we will show results of numerical calculations performed using appropriate 4-dimensional Hamiltonians (note that the nontrivial part of the Hamiltonian (\ref{eq:Heegg}) acts in a 4-dimensional subspace), but these results will be discussed with the help of generally very accurate analytical formulas obtained using the Hamiltonians from Eq.~(\ref{eq:Hk_ud}).  

\section{Effects of nonadiabatic charge dynamics on singlet return probability} \label{sec:nonadiabatic}
During a perfectly adiabatic sweep of detuning from $(4,0)$ to $(3,1)$ charge regimes the $\ket{S_{eg}}$ state should be created if $t_{eg}$ is finite - and if we have $t_{eg}\! =\!0$, then $t_{ee}$ and $t_{gg}$ have to be finite, and either $\ket{S_{gg}}$ or $\ket{S_{ee}}$ will be created, depending in which of the dots the valley splitting is larger. However, experiments on charge separation and subsequent shuttling of $R$ dot in Si/SiGe structures in \cite{Struck_NC24} showed signatures of initialization of more than one $S(3,1)$ state, as the singlet return probability signals for $R$ dot shuttled for distance $d$ and back contained
 clear signatures of presence of two frequencies, corresponding to two distinct spatial dependencies of $g$-factors of the shuttled qubit. In this Section we focus on the possible reasons for initialization of mixture of two (or more) singlets, and its consequences for the singlet return probability signal. 

\subsection{Mechanisms leading to nonadiabaticity during charge separation} \label{sec:nonadiabatic_mechanisms}
\begin{figure}[tb]
\includegraphics[width=\columnwidth]{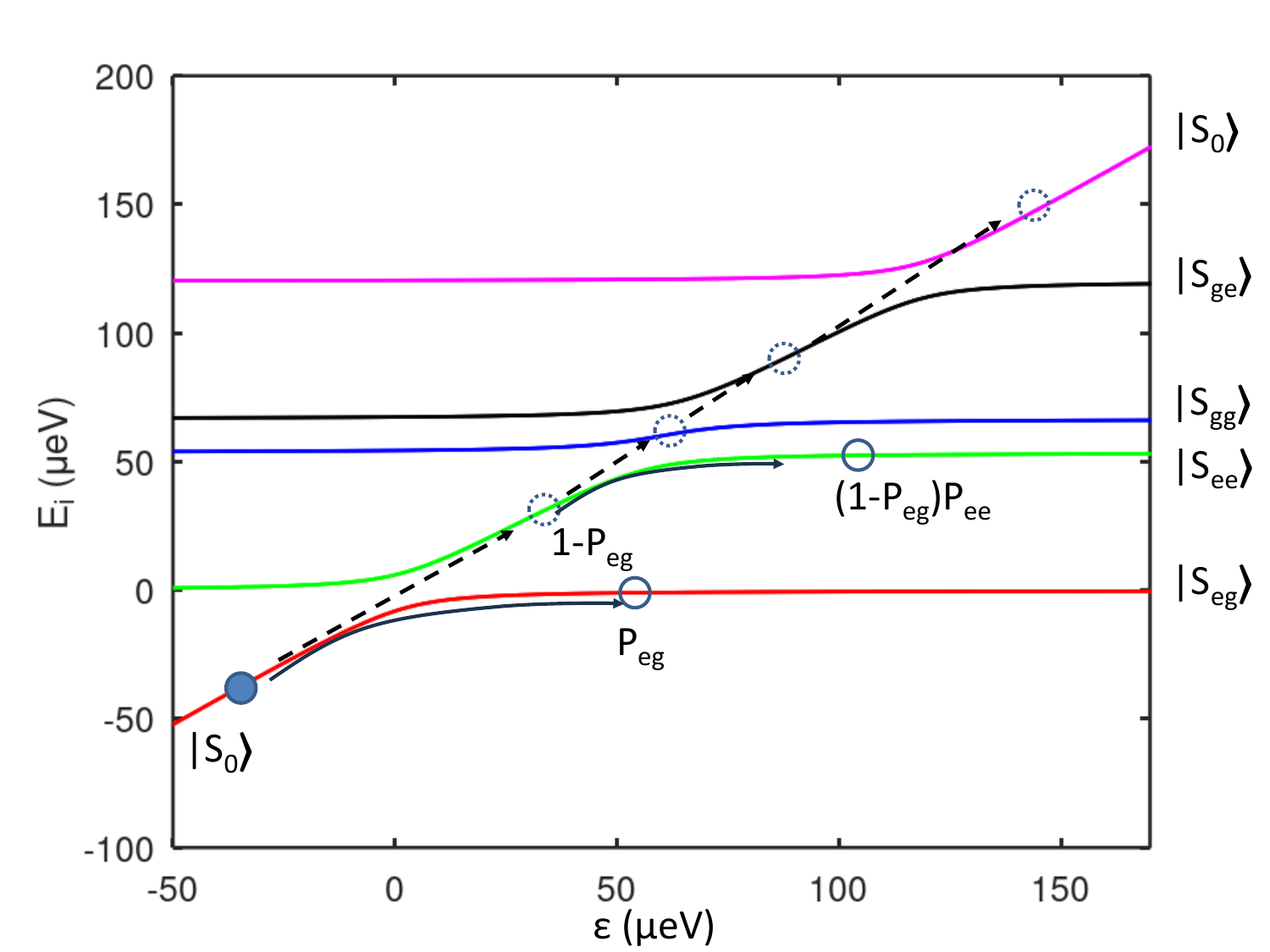}
    \caption{Solid lines: spectrum of Hamiltonian (\ref{eq:Hsinglets}) as a function of detuning $\epsilon$ for $\EVD{R} \! = \! 53.5$ $\mu$eV, $\EVD{L} \! =\! 66.6$ $\mu$eV, and all tunnel couplings equal to $t_c\! =\! 10$ $\mu$eV. $\ket{S_0}$ is the singlet ground state of the $(4,0)$ charge configuration, while $\ket{S_{\nu\mu}}$ are the 4 singlets with $(3,1)$ charge configuration, with valley occupations depicted in Fig.~\ref{fig:labelling}. Solid (dashed) lines with arrows show adiabatic (diabatic) evolution pathways: system initialized in $\ket{S_0}$ at large negative detuning upon detuning sweep evolves to $\ket{S_{eg}}$ state with probability $P_{eg}$, while it stays in $\ket{S_0}$ with probability $Q_{eg}\! \equiv \! 1-P_{eg}$, and subsequently it can adiabatically evolve into $\ket{S_{ee}}$, with the probability of populating that state being $(1-P_{eg})P_{ee}$, etc. Pathways of subsequent evolution, with exception of the ``fully diabatic'' one for which the system stays in $\ket{S_0}$ state at the end of detuning sweep, and probabilities of reaching other states, are not shown for clarity of the Figure. 
    }
		\label{fig:nonadiabatic}
\end{figure}

We assume that the system is initialized in the $S(4,0)$ state at negative detuning.  When $\epsilon$ is then changed very slowly to a positive value $\epsilon \! \gg \! t_{eg}$, so that the dynamics is truly adiabatic, the state reached in the $(3,1)$ regime is the $\ket{S_{eg}}$ singlet. More precisely, the modulus squared of overlap of the final state with $\ket{S_{eg}}$ is given by $P_{eg}\! \equiv \! 1- Q_{eg} \! \approx \! 1$, where the small probability electron not being transferred to the $R$ dot at the first anticrossing in Figs.~\ref{fig:all_funnels} and \ref{fig:nonadiabatic} is given by the Landau-Zener formula \cite{Shevchenko_PR10},
\beq
Q_{eg} =  \exp\left(-4\pi |t_{eg}|^2 / \hbar v_\epsilon \right)   \label{eq:QLZ}
\eeq
where $v_\epsilon$ is the detuning sweep rate, i.e.~$\epsilon(t) \propto v_\epsilon t$ near the anticrossing. Now it is clear that for the sweep to be ``slow enough'' to adiabatically create the $\ket{S_{eg}}$ state we need $v_{\epsilon} \! \ll \!  4\pi |t_{eg}|^2/\hbar$. For $|t_{eg}|\! = \! 10$ $\mu$eV this means $v_{\epsilon} \! \ll \! 2000$ $\mu$eV/ns, but for $|t_{eg}|\! = \! 1$ $\mu$eV  this means $v_{\epsilon} \! \ll \! 20$ $\mu$eV/ns. In \cite{Volmer_NPJQI24} the estimated detuning sweep rate was in the $v_{\epsilon}\! \in \! [500,2000]$ $\mu$eV/ns range,
so an assumption of fully adiabatic charge separation might not be justified: 
if $v_{\epsilon}\! =\! 2000$ $\mu$eV then $Q_{eg} \!< \! 0.1$ only when $|t_{eg}|\! > \! 15 $ $\mu$eV. 

Another possible cause of failure of adiabatic charge separation, which in fact becomes more relevant as the detuning sweep rate is decreased, is  high-frequency charge noise in detuning \cite{Krzywda_PRB20,Krzywda_PRB21}. When $|t_{\mu\nu}| \! \lesssim \! 10$ $\mu$eV, fluctuations in detuning that have appreciable spectral density at frequencies in  GHz range can lead to excitation of the system out of its instantaneous ground state to a higher energy state when the two are mixed by tunnel coupling near the tunnel-induced anticrossing. Let us denote the corresponding excitation rate at the anticrossing of $\ket{S_0}$ with $\ket{S_{\nu\mu}}$ by $\Gamma^{\nu\mu}_+$.
As discussed in \cite{Krzywda_PRB21}, the resulting probability of ``incoherent'' (i.e.~detuning noise-induced) failure of charge separation at given anticrossing, scales as $\Gamma^{\nu\mu}_{+}/v_\epsilon$, as slower sweep leads to longer time spent at the anticrossing. When the tunneling induced gap $2t_{\nu\mu}$ is larger than $k_{B}T$, the excitations are exponentially suppressed, $\Gamma_{+}^{\nu\mu}\! \propto \exp(-2|t_{\mu\nu}|/k_{B}T)$. However, for typical temperature $T\! =\! 100$ mK and $|t_{\mu\nu}| \! \leq\! 10$ $\mu$eV we have $2|t_{\mu\nu}|/k_{B}T \! \leq\! 2$, and such a suppression is rather weak. For such low tunnel couplings high-frequency charge noise can thus become quite efficient at introducing charge separation errors during the detuning sweep. 

After a failure of adiabatic charge separation at the first encountered anticrossing, the $(4,0)$ state can become converted into $(3,1)$ state at the subsequently encountered tunnel-induced anticrossing of singlets, see Fig.~\ref{fig:nonadiabatic} for an illustration. With $\epsilon$ varying linearly with time, the Hamiltonian from Eq.~(\ref{eq:Hsinglets}) corresponds to a 5-level generalization of the Landau-Zener problem, with energy of $\ket{S_0}$ state crossing through energies of four $(3,1)$ singlet states. Fortunately, in the relevant here case of a single level crossing multiple levels, the unperturbed energies of which are parallel lines as functions of $\epsilon$, there exists an exact solution for the occupations of all levels after the end of the detuning sweep \cite{Brundobler_JPA93}. Labeling here the $(3,1)$ singlets with $k\! =\! 1 \ldots 4$ index in the order in which the anticrossings with $\ket{S_0}$ state occur, the probability of creation of $k$-th charge separated singlet, $p^\infty_k$ by sweep of detuning to large positive values is given by
\beq
p^\infty_{k} = (1-Q_{k}) \prod_{n=1}^{k-1}Q_n  \,\, , \label{eq:pinfk}
\eeq 
where $Q_k$ is given by Eq.~(\ref{eq:QLZ}) with an appropriate $t_k$, and the probability of failure of charge separation, i.e.~the occupation of the $(4,0)$ state (the $\ket{S_0}$ singlet), is given by
\beq
p^\infty_0 =  \prod_{n=1}^{4}Q_n  \,\, . \label{eq:pinf0}
\eeq

\subsection{Contributions to the singlet return probability signal} \label{sec:PS_decomposition}
If we do not assume that the detuning sweep is adiabatic, we have to contend with a multi-level Landau-Zener problem defined by Hamiltonian \eqref{eq:Hsinglets} with $\epsilon(t) \propto v_{\epsilon}t$. At the end of the sweep from negative to positive $\epsilon$, the density matrix of the system is given by
\beq
\hat{\rho}(0) = \sum_{k=0}^{4} p^\infty_k \ket{S_k}\bra{S_k} + \sum_{k\neq k'} c_{k,k'}\ket{S_k}\bra{S_{k'}} \,\, ,
\eeq
where $k=0,\ldots,4$ labels all the singlets, see Eq.~(\ref{eq:Hsinglets}), $p^{\infty}_k$ are the final occupations of all the singlet states (with $p_0^\infty$ being the probability of {\it failure} of charge separation), and $c_{k,k'}$ are the coherences between the singlets. Subsequent evolution due to the total Hamiltonian (which is a direct sum of $\hat{H}_{\mu\nu}$ Hamiltonians from Eqs.~(\ref{eq:Heg}), (\ref{eq:Hge}), and (\ref{eq:Heegg})), 
 leaves $p_k^\infty$ unchanged, while the coherences become time-dependent. 

We now assume that these $c_{k,k'}(t)$ dephase to zero on the timescale of nontrivial dynamics and decay of $P_S(t)$ signal. This is a very natural assumption for $c_{0,k} = c_{k,0}^*$ coherences between states with distinct charge distribution, as these decay due to charge-noise induced fluctuations of energy levels in the two quantum dots (for survey of reported magnitudes of charge noise in Si/SiGe QDs see \cite{Kranz_AM20}), and the characteristic timescale of decay of these coherences is $\lesssim 1$ ns. 
As for $c_{k,k'}$ coherence with $k,k' \! \geq \! 1$, they will dephase due to fluctuations of valley splittings, but also due to the following mechanism specific to the way in which they are created. In experiments involving charge separation of spin singlet ans subsequent PSB measurement, the detuning is swept from $\epsilon_i \! \ll \! -|t|$ to $\epsilon_f \! \gg \! |t|$, the systems is kept at this $\epsilon_f$ for waiting time $t_w$, and then the detuning is swept back to $\epsilon_i$. When two $S(3,1)$ states that differ in energy of $E_V$ (equal to valley splitting in one of the dots, or the sum of these two valley splittings) are populated due to nonadiabatic character of dynamics, the total time spent in the superposition of these states is $t_w + 2(\epsilon_f-E_{V})/v_{\epsilon}$.
In a realistically noisy system, the rms of quasistatic shift of detuning due to charge noise is $\sigma_\epsilon \! \sim \! 5$ $\mu$eV (corresponding to coherence time of DQD charge qubit of $T_{2,c}^* \! =\! \hbar\sqrt{2}/\sigma_\epsilon \! \sim  \! 0.2$ ns). When both $\epsilon_0$ and $\epsilon_f$ are shifted by $\delta \epsilon \! \sim\! \sigma_\epsilon$, the time spent in the superposition of the two singlets changes by $\delta t \! \sim \! 2\sigma_\epsilon/v_{\epsilon}$. The resulting rms of phase of the relevant $c_{k,k'}$ is $\sigma_\phi \! \approx \! E_{V}\delta t /\hbar$. For $E_{V} \! \approx \! 50$ $\mu$eV, $\sigma_\epsilon \! =\! 5$ $\mu$eV, and the range of $v_{\epsilon}\! =\! 500-2000$ $\mu$eV/ns used in \cite{Struck_NC24,Volmer_NPJQI24}, we have $\delta t \! \sim \! 10$ ps that gives $\sigma_\phi \! \sim \! 1$. 
The dephasing due to this mechanism is thus significant, justifying our neglect of coherences $c_{k,k'}$ even at shortest values of $t_w$ considered in experiments. Consequently, we assume that at time $t$ at which the backward sweep of detuning to the PSB regime is initiated the state of the system is given by
\beq
\hat{\rho}(t) =p_0^\infty\ket{S_0}\bra{S_0} +  \sum_{\mu,\nu=e,g} p^\infty_{\mu\nu} \meanqs{\ket{\Psi_{\mu\nu}(t)}\bra{\Psi_{\mu\nu}(t)} }  \,\, ,\label{eq:rho_sep}
\eeq
in which we have switched back to labeling the $(3,1)$ singlets with $\mu\nu$ indices describing the valley occupation pattern. In this expression the overbar denotes averaging over distributions of all the quasi-statically fluctuating parameters in the Hamiltonian that governs the evolution from the initial $\ket{S_{\mu\nu}}$ states to the final $\ket{\Psi_{\mu\nu}(t)}$ states. The latter can be written as
\begin{align}
\ket{\Psi_{\mu\nu}(t)} & = \exp(-i\hat{H}t/\hbar) \ket{S_{\mu\nu}} \nonumber\\
& = a_{\mu\nu}(t)\ket{S_{\mu\nu}} + \sum_{m=-1,0,1} b_{\mu\nu}^{m}(t)\ket{T_{\alpha\beta}^m}  \,\, , \label{eq:Psi_k}
\end{align}
with $T_{\alpha\beta}^m$ being the appropriate triplet states coupled to $\ket{S_{\mu\nu}}$ by $\hat{H}$, see Table~\ref{tab:resonances}. As discussed in Sec.~\ref{sec:H_singlets}, instead of using the Hamiltonian acting in 17-dimensional space spanned by $\ket{S_0}$ and 16 states from $(3,1)$ charge configuration, one can use Hamiltonians from Eqs.~(\ref{eq:Heg}), (\ref{eq:Hge}), and (\ref{eq:Heegg}), or even a Hamiltonian $\hat{H}_{\mu\nu,\alpha\beta}$ from Eq.~(\ref{eq:Hk_ud}) with appropriately chosen values of subscript indices, in order to model the time dependence of state evolving from a given $\ket{S_{k}}$ state.

If the sweep of detuning back to $\epsilon_i$, at which the PSB readout will take place, is done adiabatically with respect to tunneling, i.e.~with $v_{\epsilon} \! \ll \! 4\pi |t_{\mu\nu}|^2/\hbar$ for every $t_{\mu\nu}$, each $\ket{\Psi_{\mu\nu}}$ gets converted to $\ket{\Psi^{a}_{\mu\nu}}$ in which $\ket{S_{\mu\nu}}$ is replaced by $\ket{S_0}$, so that the final state $\hat{\rho}^a(t)$ is given by Eq.~(\ref{eq:rho_sep}) in which each $\ket{\Psi_{\mu\nu}}$ is replaced by such $\ket{\Psi^{a}_{\mu\nu}}$.
The singlet return probability is then given by
\begin{align}
P_{S}^{(a)}(t) & \equiv \mathrm{Tr}\Big( \ket{S_0}\bra{S_0} \hat{\rho}^a(t) \Big ) \,\, , \nonumber\\ 
& = p_0^\infty + \sum_{\mu,\nu} p_{\mu\nu}^\infty \meanqs{|a_{\mu\nu}(t)|^2} \,\, \nonumber\\
& \equiv p_0^\infty + \sum_{\mu,\nu} p_{\mu\nu}^\infty P_{S,{\mu\nu}}(t), \label{eq:PSa}
\end{align}
in which we have identified $ \meanqs{|a_{\mu\nu}(t)|^2}$ with singlet return probability signal, $P_{S,{\mu\nu}}(t)$, that would be observed when $\ket{S_k}$ state is prepared by charge separation, and then PSB measurement is done after a slow detuning sweep that adiabatically converts $\ket{S_k}$ state into the $\ket{S_0}$ state. The calculation of $P_{S,k}(t)$ when fluctations of parameters in the Hamiltonian is disregarded is  detailed in Sec.~\ref{sec:ST_oscillations}, while in Sec.~\ref{sec:dephasing} the noise-averaged quantity $\meanqs{|a_{\mu\nu}(t)|^2}$ (i.e.~the $P_{S,{\mu\nu}}(t)$ signal including the effects of dephasing due to quasi-static fluctuations in spin and valley splitting) is calculated.

However, it is more natural to assume that the sweep of detuning back to $\epsilon_i$ is a time-reversal of the sweep that resulted in the imperfect charge separation that resulted in state from Eq.~(\ref{eq:rho_sep}). In such a case, in each $\ket{\Psi_{\mu\nu}}\bra{\Psi_{\mu\nu}}$ density operator in the $\hat{\rho}^{na}(t)$ state obtained by non-adiabatic ($na$) sweep, the occupation  of $\ket{S_0}$ state is given by the occupation of $\ket{S_{\mu\nu}}$ state before the sweep (i.e.~$|a_{\mu\nu}(t)|^2)$), multiplied by the probability of converting this state into $\ket{S_0}$. For the time-reversed sweep, this probability is equal to the probability of obtaining $\ket{S_{\mu\nu}}$ state from $\ket{S_0}$ state by the charge-separating sweep, i.e.~it is given by $p_{\mu\nu}^\infty$. If the dynamics of diagonal elements of $\hat{\rho}^{(na)}$ is then frozen during the PSB readout, the measured singlet return probability is
\beq
P_{S}^{(na)}(t) = \left(p_0^\infty\right)^2 + \sum_{\mu,\nu} \left(p_{\mu\nu}^\infty\right)^2 \,\, P_{S,{\mu\nu}}(t) .\label{eq:PSna}
\eeq
It should be stressed that it is by no means obvious that the diagonal elements of $\hat{\rho}^{(na)}$ are indeed frozen during the readout. In fact, in \cite{Struck_NC24,Volmer_NPJQI24} after return sweep of detuning to the PSB range of values, the finite interdot tunnel coupling was maintained for 500 ns before the charge state was frozen in by reducing the tunnel coupling, and the charge state readout by the nearby SET is begun. During this 500 ns long stage, inelastic tunneling from $S(3,1)$ states to the $S_0$ state is possible. 
In the simplest scenario in which these inelastic tunneling processes occur after sweeping the detuning to the PSB regime so quickly that any dynamics of $\hat{\rho}^{na}$ except for $S(3,1) \rightarrow S(4,0)$ conversion can be neglected. The resulting singlet return probability signal is then given by Eq.~(\ref{eq:PSa}) - all the nonadiabatic effects that occurred during the sweep of detuning back to $\epsilon_i$ are now simply ``healed'' by inelastic tunneling. 

\begin{figure}[tb]
\includegraphics[width=\columnwidth]{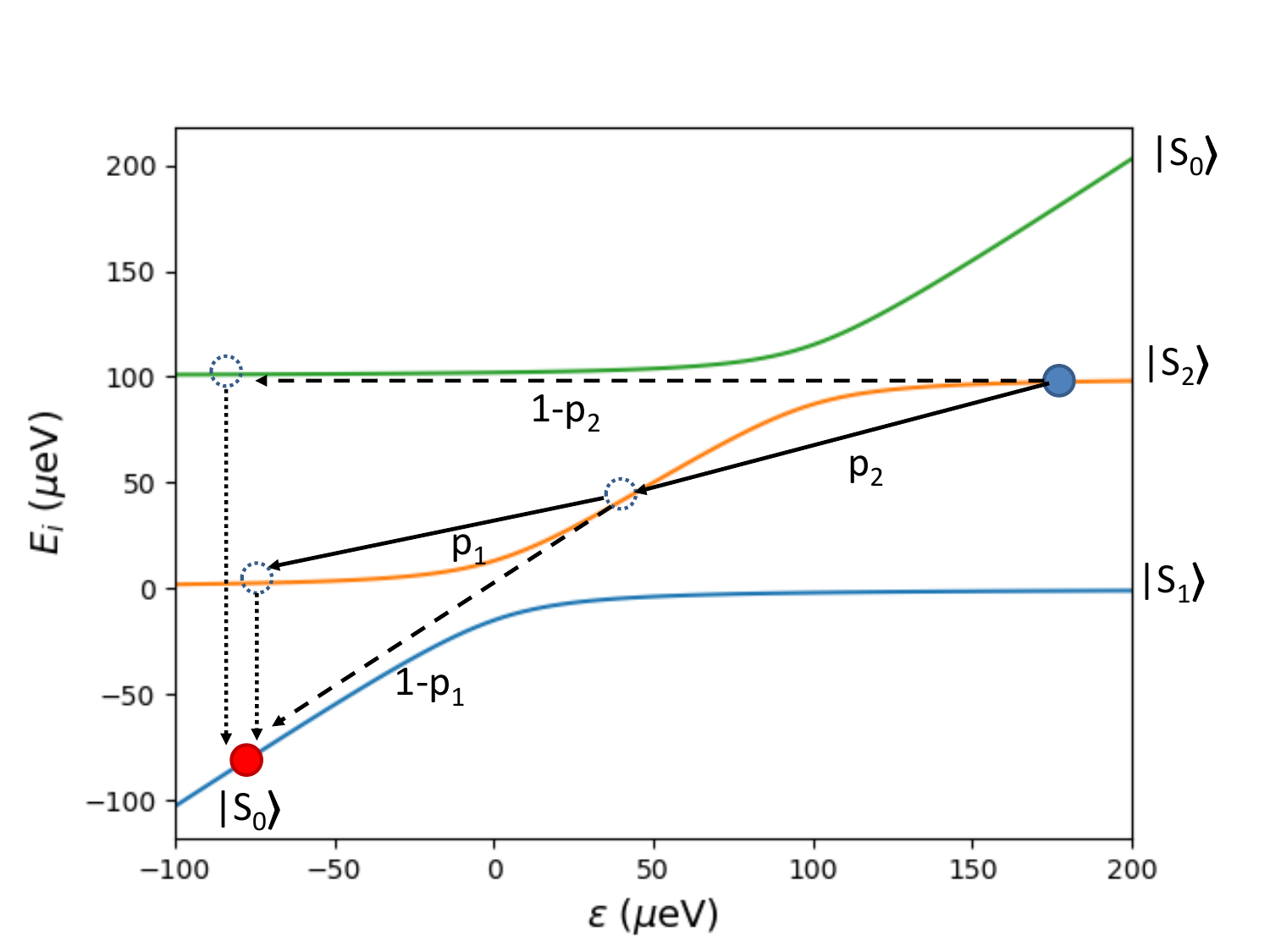}
    \caption{Illustration of  pathways of getting back from a given $S(3,1)$ state to $\ket{S_0}$ which is the singlet ground state of the $(4,0)$ charge configuration. Solid lines show the spectrum of Hamiltonian for a system with only two $S(3,1)$ singlets (labeled as $\ket{S_1}$ and $\ket{S_2}$) as a function of detuning $\epsilon$ for $\EVD{} \! = \! 100$ $\mu$eV, and both tunnel couplings equal to $t_c\! =\! 10$ $\mu$eV.  Initial state at large positive detuning is $\ket{S_2}$ (filled blue circle), while the desired final state in which PSB measurement works as intended is $\ket{S_0}$ at negative detuning (filled red circle). Lines with arrows correspond to respective pathways of dynamcis: adiabatic evolution during detuning sweep back to negative values (solid), diabatic evolution during this sweep (dashed), and inelastic interdot tunneling in PSB regime before raising of the tunnel barrier (points). $p_k$ is the probability of adiabatic passage through anticrossing of $\ket{S_k}$ with $\ket{S_0}$. 
    Without the inelastic processes the probability to obtain the signal corresponding to singlet in the PSB measurement when starting from $\ket{S_2}$ occupied at large positive detuning is given by $p_2(1-p_1)$. In order to reach the $\ket{S_0}$ state with unit probability inelastic tunneling must occur on timescale shorter than the time of waiting in $(4,0)$ configuration with tunnel barrier height allowing for interdot coupling.
    }
		\label{fig:way_back}
\end{figure}

Let us illustrate the above with an example involving only the first two $(3,1)$ singlets --- $S_{eg}$ and $S_{ee}$, labeled here $S_1$ and $S_2$ --- encountered during the non-adiabatic detuning sweep from $\epsilon_i$ to $\epsilon_f$ and then back. 
During the forward detuning sweep, the first of them, $S_{eg}$, is initialized with probability $p_1 \! =\! 1-Q_1$, where $Q_1$ is given by Eq.~(\ref{eq:QLZ}). The probability of diabatic passage through the first anticrossing is $Q_1$, and since the probabilities of subsequent adiabatic (diabatic) passage through the second anticrossing are $p_2$ ($Q_2 \! = \! 1-p_2$), the occupation of $S_2$ state at detuning $\epsilon_f$ is given by $p_2 (1-p_1)$, in agreement with Eq.~(\ref{eq:pinfk}), while the occupation of $S_0$ state (the probability of failure of charge separation) is $Q_1 Q_2 \! =\! (1-p_1)(1-p_2) \! = \! p^\infty_0$, in agreement with Eq.~(\ref{eq:pinf0}). After evolution for time $t$, the amplitudes of having $S_{1(2)}$ states are $a_{1(2)}(t)$, see Eq.~(\ref{eq:Psi_k}). Next, during the return sweep of detuning, the $S_1$ state is adiabatically converted into $S_0$ with probability $p_1$, leading to $(p_1)^2 \meanqs{|a_1(t)|^2}$ contribution to the $P_S(t)$ signal, while with probability $p_1(1-p_1)$ the system is ``stuck'' in $S_1$.
Similarly, for $S_2$ to be converted in $S_0$, the state needs to follow a time-reversed path of adiabatic passage through the second anticrossing and a {\it diabatic} passage through the first one, so that the resulting contribution to $P_S(t)$ is $[p_2(1-p_1)]^2 \meanqs{|a_1(t)|^2}\! =\! (p_2^\infty)^2\meanqs{|a_1(t)|^2}$. The ``wrong'' pathways of evolution lead to final occupation $S_2$ of $(1-p_2)p^\infty_2$, and to an additional contribution of occupation of ``stuck'' $S_1$ state given by $p_2 p_1 p^\infty_2$.
The processes that occur for evolution starting from $S_2$ in $(3,1)$ regime are illustrated in Fig.~(\ref{fig:way_back}).
Finally, the $S_0$ remains occupied after the backwards sweep with probability $(p_0^\infty)^2$, while its evolution during this sweep contributes $p_2 p_0^\infty$ to occupation of $S_2$ and $(1-p_2) p_1 p_0^\infty$ to occupation of $S_1$. If we now assume that the occupations of $S_1$ and $S_2$ states are rapidly converted by inelastic tunneling into occupations of $S_0$, the terms in Eq.~(\ref{eq:PSna}) are modified in the following way:
\begin{itemize}
    \item The contribution to $P_S(t)$ from not leaving the $S_0$ state during the whole dynamics changes from $(p_0^\infty)^2$ to $(p_0^\infty)^2 + p_2 p_0^\infty + (1-p_2) p_1 p_0^\infty \! =\! p_0^\infty$. 
    \item The contribution from $S_2$ state that evolved for time $t$ at $\epsilon\! =\! \epsilon_f$ changes from $(p_2^\infty)^2\meanqs{|a_2(t)|^2}$ to $[(p_2^\infty)^2 + (1- p_2) p_2^\infty + p_2 p_1 p_2^\infty]\meanqs{|a_1(t)|^2} \! =\! p_2^\infty \meanqs{|a_2(t)|^2}$.
    \item The contribution from $S_1$ state changes from $p_1^\infty  \meanqs{|a_1(t)|^2}$ to $[p_1^\infty + (1-p_1)] \meanqs{|a_1(t)|^2} \! =\! p_1^\infty  \meanqs{|a_1(t)|^2}$.
\end{itemize}
In this way $P_S(t)$ given by Eq.~(\ref{eq:PSna}) is changed to the one given by Eq.~(\ref{eq:PSa}) due to fast inelastic tunneling of the $S(3,1)$ state into the $S(4,0)$ state at the beginning of the PSB readout stage. 

\section{Oscillations of the singlet return probability signal near the spin-valley hotspot}  \label{sec:ST_oscillations}
In absence of spin-valley mixing, the time-dependence of $P_{S,\mu\nu}(t)$ defined in Eq.~(\ref{eq:PSa})  is only due to $S$-$T_0$ mixing by the Zeeman splitting difference $\DEZ{\mu\nu}$. After averaging the result over a quasi-static distribution of spin splittings we obtain the well-known result \cite{Taylor_PRB07},
\beq
P_{S,\mu\nu} \! = \! \frac{1}{2} +\frac{1}{2}\cos\left(\DEZ{\mu\nu}t \right) \exp(-\frac{1}{2}\sigma^2_{\mu\nu} t^2) \,\, , \label{eq:PS_nomixing}
\eeq 
in which $\sigma_{\mu\nu}$ is the rms of distribution of $\DEZ{\mu\nu}$.  We will now include the effects of spin-valley mixing on the $P_{S,\mu\nu}(t)$ signal when the Zeeman splitting is close to the valley splitting in one of the dots.

Depending on the kind of singlet $\ket{S_{\mu\nu}}$ and $\AEZ{} \! \approx \! \EVD{L}$ or $\EVD{R}$, we are thus dealing with one of $8$ resonances from Table \ref{tab:resonances}. We will use  the number of resonance  $r\! =\! 1 \ldots 8$ from this Table as an index identifying the relevant parameter, i.e.~$S_r$ will stand for respective $S_{\mu\nu}$, $T_r$ will stand for the respective $T^p_{\alpha\beta}$, $\hat{H}_r$ will stand for the respective $\hat{H}_{\mu\nu,\alpha\beta}$ from Eq.~(\ref{eq:Hk_ud}), etc.
The probability amplitude of the system initialized in $\ket{S_r}$ state to be in this state at time $t$ is $a_r(t) \! =\! \bra{S_r} e^{-i\hat{H}_r t}\ket{S_r}$, and the corresponding singlet return probability is $P_{S,r}(t) \! \equiv \! |a_r(t)|^2$.

\begin{figure}[tb!]
    \centering
        \includegraphics[width=0.8\columnwidth]{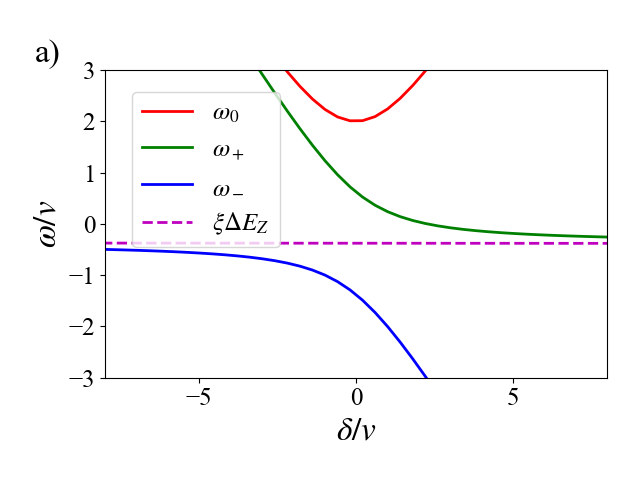}
        \includegraphics[width=0.8\columnwidth]{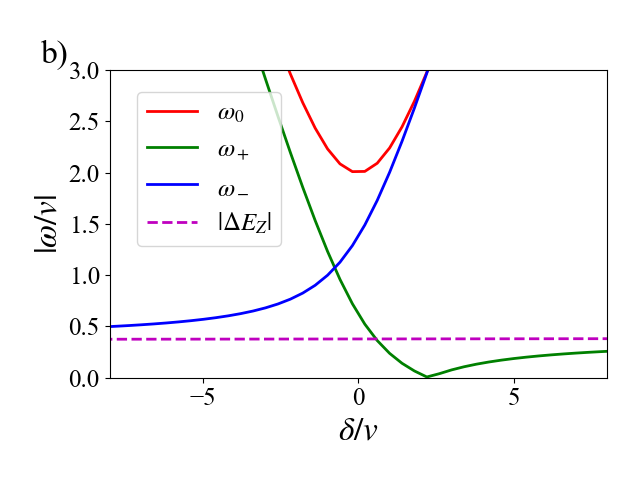}
        \includegraphics[width=0.8\columnwidth]{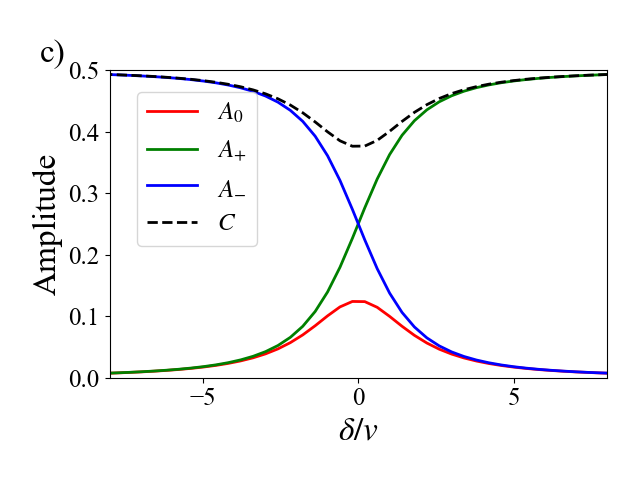}
    \caption{a) Frequencies present in $P_S(t)$ signal (thick solid lines) in the vicinity of spin-valley resonance in one of the dots, given by Eqs.~\eqref{eq:w0} and \eqref{eq:wpm}, and $\xi_r\DEZ{r}$ frequency (dashed line) that would be observed in absence of spin-valley coupling. The calculation is done for $\xi\! =\! 1$, $p\! =\! -1$, and $\Delta E_Z \! < \! 0$. The frequencies and detuning $\delta$ are expressed in units of $v$, the spin-valley coupling at a relevant resonance, making the features of the plots other than the position of the dashed no-spin-valley-coupling line universal. The used value of $|\Delta E_Z|$ at resonance corresponds to $-0.38 v$. This is realized, for example, for resonance involving $\ket{S_{eg}}$ and electron in the left dot, when $\EVD{L} \! =\! 66.6$ $\mu$eV and $v\! =\! v_L\! = 58$\,neV, which are the values measured for a DQD in \cite{Volmer_NPJQI24}. For these values, the span of $\delta/v$ shown in the Figure corresponds to a span of 8\,mT around an anticrossing at $\approx 570$\,mT. 
    b) The same, but with absolute value of the frequencies plotted. c) The corresponding amplitudes from Eq.~\eqref{eq:CA}.
    }
    \label{fig:omegaA}
\end{figure}

Let us introduce the detuning from the relevant spin-valley anticrossing,
\beq
\delta_{r} = \frac{1}{2}\xi_{r}\DEZ{r} + p_r(\EVD{r} - \AEZ{r})  \,\, , \label{eq:delta}
\eeq
and the parameter $\gamma_r$ defined as
\beq
\gamma_{r} = \frac{1}{2}\xi_{r}\DEZ{r} - p_r(\EVD{r} - \AEZ{r}) \,\, , \label{eq:gamma}
\eeq
so that the part of Hamiltonian from Eq.~(\ref{eq:Hk_ud}) that is acting nontrivially in the relevant two-dimensional subspace is given by $\frac{1}{2}\mathds{1}\gamma_r + v\hat{\sigma}_x + \frac{1}{2}\delta_r \hat{\sigma}_z$. We also introduce the mixing angle $\theta_{r}$ given by
\beq
\cos \theta_{r} \equiv \frac{\delta_{r}}{\sqrt{\delta^2_{k} + 4|v_{k}|^2}} \,\,  , \label{eq:costheta}
\eeq
so that for $\delta_r/|v_r| \! \rightarrow \! \infty$ we have $\theta_r \! \rightarrow \! 0$, while on the other side of the anticrossing, for  $\delta_r/|v_r| \! \rightarrow \! -\infty$  we have $\theta_r \! \rightarrow \! \pi$.
\optional{Note that $\partial \delta/\partial B \!= \! (2\eta+\xi\Delta g/3)\mu_B \approx \! 116\eta$ neV/mT. This means that for $v\! \approx \! 50$ neV we have $\delta\approx 2v$ for about 1 mT detuning from the field at which the resonance occurs.} 

Using the above we can derive 
\beq
P_{S,r}(t) = C_{r} + \sum_{b=0,\pm} A_{r}^b \cos \omega_{r}^b  \,\, . \label{eq:PSbare}
\eeq
where the constant part of the signal
is $C_{r}$, and the $A_r^b$ the amplitudes of the oscillating parts are
\begin{align}
C_r & = \frac{1}{2} - \frac{1}{8}\sin^2\theta_{r} \,\, , \nonumber\\
A_{r}^0 &= \frac{1}{8} \sin^2 \theta_{r} \,\, , \nonumber\\
A_{r}^{\pm} & = \frac{1}{4}(1\pm \cos\theta_{r} )  \,\, , \label{eq:CA}
\end{align}
and the frequencies are
\begin{align}
\omega_r^0 & = \sqrt{\delta^2_{r} + 4|v_{r}|^2} \,\, , \label{eq:w0} \\
\omega_{r}^{\pm} & = \frac{1}{2}(\xi_{r}\DEZ{r} + \gamma_{r} \pm \sqrt{\delta_{r}^2 + 4|v_{r}|^2} ) \nonumber\\
& = \frac{1}{2}(2\gamma_{r} +\delta_{r} \pm  \sqrt{\delta_{r}^2 + |v_{r}|^2}) \,\, .\label{eq:wpm}
\end{align}

Let us define $s \! \equiv \! \mathrm{sgn}(\delta_r)$.
Far away from the hotspot,  i.e.~for $|\delta_r|  \! \gg \! 2 |v_r|$, after using the fact that $\gamma_{r}+\delta_{r} \! =\! \xi_{r}\DEZ{r}$, we see that only the oscillation with frequency $\omega^{s}_r \!  \approx \! \xi_r\DEZ{r}$ has appreciable amplitude given by $A^s_r \! \approx \! 1/2$. To be precise, on the low (high) magnetic field side of the resonance  we have $s\! =\! p_r$ ($s\! =\! -p_r$) and 
\beq
\omega^{s}_r   \approx  \xi_{r}\DEZ{r} +s|v_{r}|^2/|\delta_{r}| \,\, , \label{eq:wdom_off_resonant}
\eeq
while $A^{s}_r \! \approx \! \frac{1}{2} - \frac{1}{2}|v_{r}/\delta_{r}|^2$ and the other two oscillations have amplitudes are $\propto  \! \frac{1}{2} |v_{r}/\delta_{r}|^2 \! \ll \! 1$. 
The result for no spin-valley mixing from Eq.~(\ref{eq:PS_nomixing}) is thus recovered at sufficiently large detuning from the resonance. 

The behavior of the ``dominant'' frequency as the relevant resonance is approached by changing the $B$ field can be understood in the following way. For given $r$, depending on the value $\xi_r$, only one of the $\ket{\ud}$ or $\ket{\du}$ levels is coupled to $\ket{T_{r}}$, see the discussion below Eq.~(\ref{eq:Hk_ud}). As the magnetic field is varied, the energy of the $\ket{T_{r}}$ state is approaching the range of energies of $\ket{\ud}$ and $\ket{\du}$ states either from above, or below, depending on the sign of $p_r$. Consequently, the sign of perturbative renormalization of splitting between $\ket{\ud}$ or $\ket{\du}$ levels has opposite signs on the low- and high-field sides of the resonance.

\begin{figure}[tb]
\includegraphics[width=0.75\columnwidth]{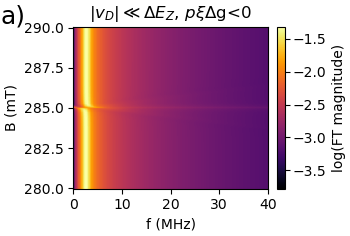}
\includegraphics[width=0.75\columnwidth]{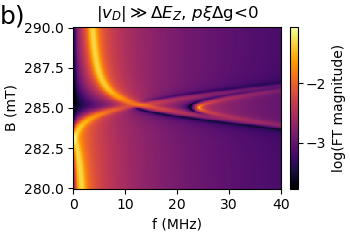}
\includegraphics[width=0.75\columnwidth]{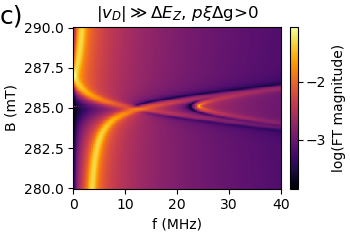}
    \caption{Fourier transform of the singlet return probability near a spin-valley anticrossing. The calculation is done for $B$ field in the vicinity of spin-valley hotspot in the $R$ dot (resonances with odd $r$ in Table \ref{tab:resonances}) for $\EVD{R}\! =\! 33$ $\mu$eV, $|\Delta g| \! =\! 6.58\cdot 10^{-4}$, and dephasing parameters $\sigma_{ZR}\! =\! 1$ neV, $\sigma_{VR} \! =\! 10$ neV.  Zero frequency components have been removed by subtracting the long-time value of $P_S(t)$ from the signals before Fourier-transforming them.
		(a) Weak coupling results for $v_D \! =\! 5$ neV, for which $|v_D|\! \approx \! |\DEZ{}|/2$. 
    (b,c) Two possible patterns of frequency and magnetic field dependence of Fourier transform of singlet return probability near a spin-valley anticrossing when $v_D \! =\! 50$ neV, so that $|v_D|\! \gg \! \DEZ{}$. 
    The result depends on the sign of $p_r \xi_r\DEZ{r}$ quantity. When calculations are done in the vicinity of spin-valley hotspot in the $L$ dot, the two patterns are exchanged, i.e.~the pattern from (c) appears when $p_r\xi_r \DEZ{r}\! < \! 0$ for even $r$. 
    }
		\label{fig:single}
\end{figure}

\begin{figure}[tb]
\includegraphics[width=0.75\columnwidth]{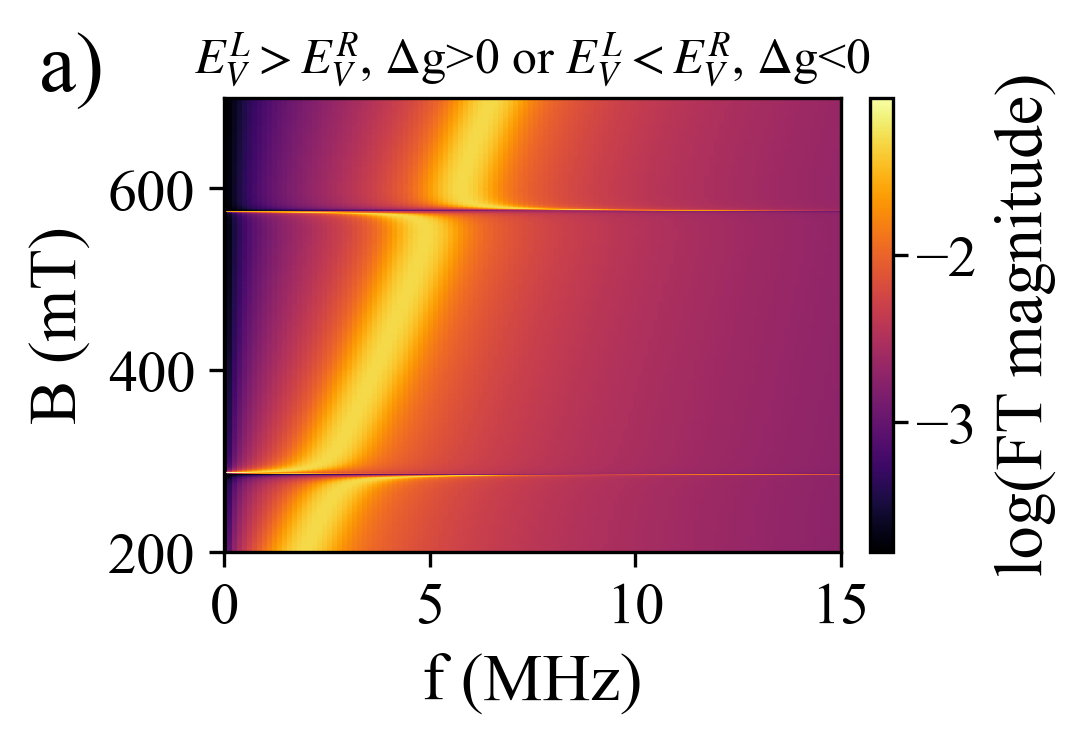}
\includegraphics[width=0.75\columnwidth]{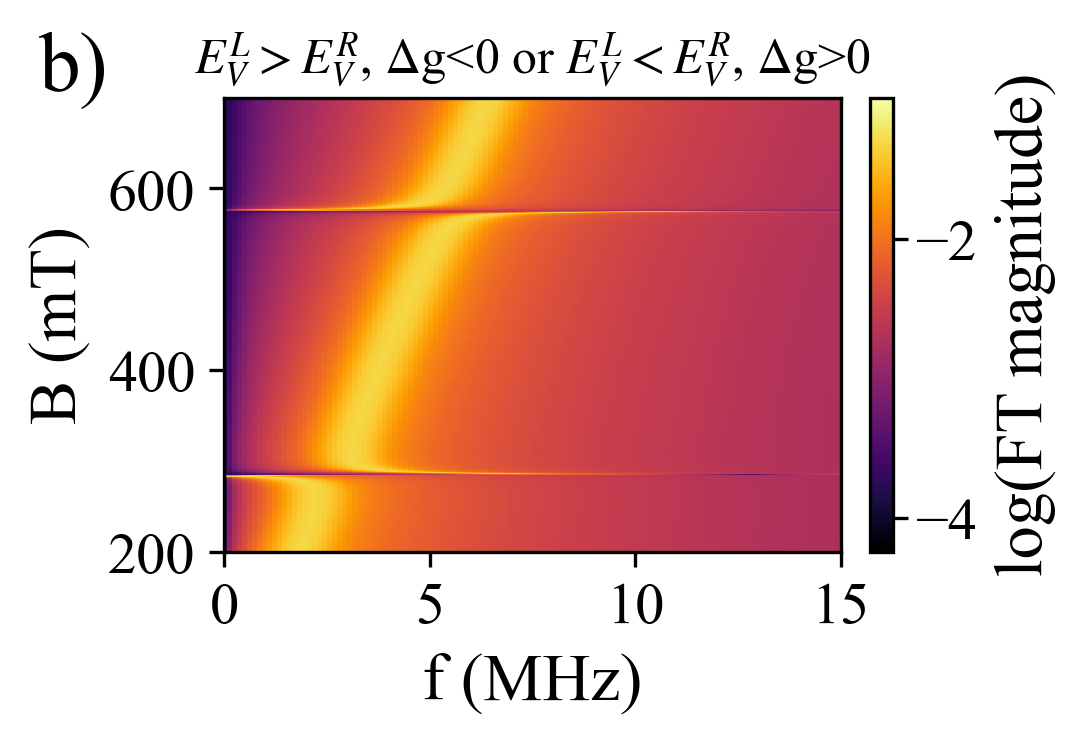}
    \caption{Two possible patterns of frequency and magnetic field dependence of the Fourier transform of singlet return probability that are realized {\it for any of the $S_{\mu\nu}$ singlets} from Fig.~\ref{fig:labelling}a depending on the signs of $\Delta g_{\mu\nu} \! =\! g_{\mu_L}-g_{\nu_R}$ and $\EVD{L}-\EVD{R}$.}
			\label{fig:both}
\end{figure}

On the other hand, close to the resonance,  when $|\delta_r|  \! \ll \! 2|v_r|$, oscillations with three distinct frequencies are present in $P_{S,r}(t)$:  we have then $A^0_r \! \approx \! 1/8$ and $A^{\pm}_r \! \approx \! \frac{1}{4} \pm \delta_r/2|v_r|$ (and note that $C_a \! \approx \! 3/8$). The frequency of the oscillation that appears only close to resonance, 
\beq
\omega_r^0 \! \approx \! 2|v_r| +  \frac{\delta^2_{r}}{4|v_r|} \,\, , \label{eq:w02nd}
\eeq
is to first order independent of $\delta_r$. 
On the other hand, the two ``dominant'' oscillations have $\omega_{\pm}$ given by
\begin{align}
\omega^{\pm}_{r} & \approx \frac{1}{2}(\xi_r\DEZ{r} + \gamma_r \pm 2|v_r| \pm \frac{\delta^2_r}{4|v_r|} )\nonumber\\
& =  \frac{3}{4}\xi_a\DEZ{a} +\frac{p_r}{2}(\AEZ{r}-\EVD{r}) \pm |v_r| \pm \frac{\delta^2_r}{8|v|} \,\, ,
\label{eq:wpmres} 
\end{align}
Three things should be noted. The first is that $\omega^{\pm}_r$ vary rapidly with changing of the magnetic field: $|\partial \omega_r^{\pm}/\partial B| \! =\! \frac{1}{2}|\partial \AEZ{r}/\partial B| \! \approx \! \mu_B$. This is half of the value of $B$ derivative of the polarized triplet energy. (Note that these frequencies become fully ``polarized triplet-like'', i.e.~$|\partial \omega_r^{\pm}/\partial B|  \! \approx \! 2\mu_B$,  only away from the resonance, when the corresponding amplitudes are very small.) 

The second observation is the appearance of the term linearly proportional to the respective valley splitting, $\EVD{r}$, in expression (\ref{eq:wpmres}).
This means that while far away from the resonance the dephasing of the signal is due to fluctuations of the ``bare'' $\DEZ{r}$ caused by interaction with nuclei, or possibly electric-noise induced fluctuations of $g$-factors, close to the resonance the valley splitting fluctuations also contribute to the dephasing of the singlet return probability signal \cite{Jock_NC22}.

The third thing to note is that exactly at resonance we have $\omega^\pm_r \! =\! \xi_r\DEZ{r} \pm |v_{r}|$. When $|v_{r}| \! \ll \! |\DEZ{r}|$, the renormalization of $\omega^{\pm}_r$ away from $\xi_{r}\DEZ{r}$ value is relatively small. On the other hand, when $|v_{r}| \! \gg \! |\DEZ{r}|$, the renormalization of oscillations having easily observable amplitudes is large, with the largest-amplitude oscillation having its frequency renormalized to zero on one side of the resonance, see Fig.~\ref{fig:omegaA}(b). This regime of large relative renormalization of singlet-triplet precession frequency near a spin-valley resonance will be our main focus, as the recent measurement \cite{Volmer_NPJQI24} of $v_D$ and $\DEZ{}$ at magnetic fields at which $\AEZ{}\! \approx \! \EVD{D}$ in Si/SiGe structures put us exactly in this regime.

Looking at a wide range of magnetic fields, the $\omega^{\pm}$ frequencies vary by orders of magnitude, see Fig.~\ref{fig:omegaA}. However, when $|\omega^{\pm}_r| \! \gg \! \DEZ{r}$,  the corresponding amplitude $A^{\pm}_r$ is very low, and such a high-frequency contribution to the total $P_S(t)$ signal is unobservable.
The $B$ field dependence of frequencies $\omega^b_r$ and amplitudes $A^b_r$ is most transparently illustrated by plotting the Fourier transform  of Eq.~(\ref{eq:PSbare}) as a function of $B$ field and frequency $f$.
In Fig.~\ref{fig:single}a we show such $f$ and $B$ dependence of Fourier transform of $P_{S,r}(t)$ when the precession frequencies are weakly renormalized, $|v_r| \! \ll \! |\DEZ{r}|$. The dominant feature of the singlet return probability signal is a slight upward or downward (depending on being on the low/high $B$ field side resonance) renormalization of the dominant frequency. 
On the other hand, when  $|v_r| \! \gg \! |\DEZ{r}|$, we obtain more interesting patterns shown in panels (b) and (c): for $v_{D} \! \approx \! 50$ neV the dominant frequency is now qualitatively renormalized in a range of $\lesssim 10$ mT width around the resonance, and an oscillation with frequency $\omega^0 \approx 2|v_r|$ appears clearly in the signal for $|\delta| \! < \! 2|v_r|$.  The frequency of high-amplitude component of $P_S(t)$ is renormalized to zero on low (high) $B$ field side of the resonance for $\xi p \Delta g<0$ ($>0$), when $|\delta_r| \! \approx \! |v_r|^2/|\DEZ{r}|$, see Fig.~\ref{fig:single}b (\ref{fig:single}c).

Careful inspection of parameters of all the resonances given in Table \ref{tab:resonances} shows that the type of pattern is controlled by the sign of $p_{r}\xi_{r}\DEZ{r}$ product. Furthermore, when we look at the range of magnetic fields covering both the spin-valley resonances, the ``global'' pattern for any of the four (3,1) singlets looks like one of the two possibilities shown in Fig.~\ref{fig:both}, depending on the sign of $\DEZ{}$ for the considered singlet and on whether we have $\EVD{L}\! > \! \EVD{R}$ or $\EVD{L}\! < \! \EVD{R}$. 

\section{Dephasing of the singlet return probability oscillations} \label{sec:dephasing} 
For a fully realistic description of the singlet return probability signal we need to include the presence of quasi-static fluctuations of spin splittings in the two dots, as they are the dominant cause of dephasing of spin qubits when no effort, such a using dynamical decoupling, is made to remove the influence of low-frequency noise on them, see Eq.~(\ref{eq:PS_nomixing}) for $P_S(t)$ away from  anticrossings, or for negligible spin-valley coupling. We label the rms of Gaussian distribution of spin splitting in dot $D$ as $\sigma_{ZD}$.
The key physical source of these fluctuations in natural Si structures is hyperfine interaction with nuclei of $^{29}$Si. In structures using isotopically enriched Si and without magnetic field gradients (in presence of which fluctuations of dot position due to electric field noise typically dominate the dephasing \cite{Yoneda_NN18,Struck_NPJQI20}), hyperfine interaction with $^{73}$Ge nuclei in the barrier material can be relevant \cite{Cvitkovich_PRAPL24}, and QD motion due to electric field noise in presence of spatial dependence of electron $g$-factor (see \cite{Volmer_NPJQI24} and \cite{Volmer_arXiv26} for examples of measured spatial dependence of $g$-factor for a shuttled Si/SiGe QD) could also contribute to $\sigma_{ZD}$ at large enough magnetic fields.

 As discussed in Sec.~\ref{sec:ST_oscillations}, near the spin-valley resonance $r$ the frequencies present in the $P_{S,r}(t)$ signal depend also on the valley splitting $\EVD{D}$ in quantum dot $D$ in which the resonance occurs. We thus introduce in the model finite rms of Gaussian distribution of valley splitting in dot $D$, $\sigma_{VD}$. It is known that $E_{V}$ depends on electric fields determining the overlap of the electron wavefunction with Si/SiGe interface and also on in-plane position of the dot \cite{Hollman_PRAPL20} (for more discussion see Sec.~\ref{sec:dephasing_exp}). Consequently, an always--present $1/f$ charge noise leads to quasi-static fluctuations of $\EVD{D}$. 

The singlet return probability in each individual measurement  is then calculated for Zeeman and valley splittings given by $\EZ{D} +\delta \EZ{D}$ and $\EVD{D}+\delta \EVD{D}$, and then an average over Gaussian distributions of all these parameters is performed in order to obtain the $P_S(t)$ signal. We assume now that the fluctuations in all the parameters are independent. This should be a very good approximations when $E_{Z}$ fluctuations are due to hyperfine interaction with the nuclear spins of $^{29}$Si and $^{73}$Ge, while $\EVD{D}$ fluctuations are due to charge noise.
We expand $\omega_r^{b}$ ($b\! =\! 0,\pm$) frequencies from Eqs.~(\ref{eq:w0}) and (\ref{eq:wpm}) that are relevant near resonance $r$ to linear order in the fluctuating parameters, and after perfoming the Gaussian averages we obtain 
\beq 
P_{S,r}(t) = C_{r} +\sum_{b=0,\pm} A^b_{r} \cos\omega^b_{r} \times e^{-(t/T_{2,r,b}^{*})^2} \,\, ,
\eeq
which is the same as Eq.~(\ref{eq:PSbare}), only with Gaussian decay envelopes multiplying the respective oscillating factors. 
The dephasing times are given by $T_{2,r,b}^* \! \equiv \! \sqrt{2}/\sigma_{r,b}$, and the variance of fluctuations of $\omega^{b}_{r}$ frequencies are given by
\begin{align}
\sigma^2_{r,0}  =& \frac{1}{4}\cos^2\theta_r \left[ (\xi_r-p_r)^2 \sigma^2_{ZL} + (\xi_r+p_r)^2 \sigma_{ZR}^2 + 4\sigma_{V,r}^2 \right] \,\, , \label{eq:sigma0} \\
\sigma^2_{r,\pm} = &\frac{1}{16} \left[ \xi_r (3\pm \cos\theta_r) + p_r (1\mp \cos\theta_r)\right]^2\sigma^2_{ZL} + \nonumber\\
& \frac{1}{16} \left[ \xi_r (3\pm \cos\theta_r) - p_r (1\mp \cos\theta_r)\right]^2\sigma^2_{ZR} + \nonumber\\
& \frac{1}{4}(1\mp \cos\theta_r)^2 \sigma^2_{V,r}\label{eq:sigmapm} \,\, ,
\end{align}
where $\sigma^2_{V,r}$ is the variance of valley splitting in dot $D$, where $D$ is taken from the $r$th row of Table \ref{tab:resonances}.

We can see that far away from the resonance, when $\cos\theta_r \! \approx \! \pm 1$, the dominant oscillation has frequency $\DEZ{r}$ and variance given by $\sigma_Z^2 \! \equiv \! \sigma^2_{ZL} + \sigma^2_{ZR}$, so it decays on timescale of $T_{2}^* \! =\! \sqrt{2}/\sigma_Z$, which is the well-known result for dephasing of singlet-triplet oscillation when spin-valley coupling is irrelevant \cite{Taylor_PRB07}, see Eq.~(\ref{eq:PS_nomixing}).
On the other hand, close to a resonance we have
\begin{align}
\sigma^2_{r,\pm} & \approx \frac{10}{16}(\sigma^2_{ZL}+\sigma^2_{ZR}) + \frac{6}{16}\xi_r p_r(\sigma^2_{ZL} - \sigma^2_{ZR}) + \frac{\sigma^2_{V,r}}{4}  + \nonumber\\
& \pm \frac{\delta_r}{16|v_r|}(\xi_r-p_r)\sigma_{ZL}^2 \pm \frac{\delta_r}{16|v_r|}(\xi_r+p_r)\sigma_{ZR}^2  \pm\frac{\delta_r}{4|v_r|}\sigma^{2}_{V,r} \,\,  ,\label{eq:sigmapmres}
\end{align}
where we have kept terms of first order in $\delta_r/|v_r|$. Inspection of Table \ref{tab:resonances} shows that for resonances involving spin-valley hotspot in L (R) dot, we have $\xi_r-p_r\! =\! 2\xi_r$ ($\xi_r-p_r\! =\! 0$), $\xi_r+p_r\! =\! 0$ ($\xi_r+p_r\! =\! 2\xi_{r}$), and $\xi_r p_r \! =\! -1$ ($\xi_r p_r \! =\! 1$), so for $L$ dot resonances, i.e.~even $r$, we have
\beq
\sigma^2_{\mathrm{even} \,\, r,\pm} \approx \frac{1}{4}\sigma^2_{ZL} + \sigma^2_{ZR} + \frac{1}{4}\sigma^2_{VL} \pm \frac{\delta_r}{8|v_L|}(\xi_r \sigma^2_{ZL} + 2\sigma^2_{VL}) \,\, , \label{eq:sigma_even}
\eeq
while for $R$ dot resonances, i.e.~odd $r$, we have
\beq
\sigma^2_{\mathrm{odd} \,\, r,\pm} \approx \sigma^2_{ZL} + \frac{1}{4}\sigma^2_{ZR} + \frac{1}{4}\sigma^2_{VR} \pm \frac{\delta_r}{8|v_R|}(\xi_r \sigma^2_{ZR} + 2\sigma^2_{VR}) \,\, . \label{eq:sigma_odd}
\eeq
If $\sigma_{VD} \! \ll \! \sigma_{ZD}$, measurement of dephasing times of the signals oscillating with $\omega^{\pm}$ frequencies at the two resonances allows for fitting of both $\sigma_{ZL}$ and $\sigma_{ZR}$. On the other hand, when $\sigma_{VD} \! \gg \! \sigma_{ZD}$, such measurement allow for fitting of values of both $\sigma_{VL}$ and $\sigma_{VR}$. 

Away from resonance, for $|\delta_{r}| \! \gg \! 2|v_{r}|$, starting from Eq.~(\ref{eq:sigmapm}) we obtain that variance of the ``dominant'' frequency $\omega^{s}_r$ (where $s\! =\! \mathrm{sgn}(\delta)$), which has amplitude $A^{s}_{r} \! \approx \! 1/2$, is given by 
\begin{align}
\sigma^2_{\mathrm{even} \,\, r}  & \approx \sigma^2_{ZL}+\sigma^2_{ZR}  - \frac{2|v_L|^2}{\delta_r^2}\sigma_{ZL}^2 + \frac{|v_L|^4}{\delta_r^4} \sigma^2_{VL} \,\, , \label{eq:sigma_off_even} \\
\sigma^2_{\mathrm{odd} \,\, r}  & \approx \sigma^2_{ZL}+\sigma^2_{ZR}  - \frac{2|v_R|^2}{\delta_r^2}\sigma_{ZR}^2 + \frac{|v_R|^4}{\delta_r^4} \sigma^2_{VR} \,\, . \label{eq:sigma_off_odd} 
\end{align}
We see that when $\sigma_{VD} \! \ll \! \sigma_{ZD}$, the dephasing time  of the singlet return probability signal approaches the value unperturbed by spin-valley coupling very rapidly once detuning becomes larger than $2|v_r|$, i.e.~once the mixing with the polarized triplet is suppressed. From Eqs.~(\ref{eq:sigma_even})-(\ref{eq:sigma_off_odd}) we see then that as long as $\sigma_{ZL}$ and $\sigma_{ZR}$ are of the same order of magnitude, the $T_{2}^{*}$ of the highest-amplitude component of the signal varies only slightly in a narrow range of $B$ fields (of $\approx 2$ mT width for $|v_r|\! \approx \! 50$ neV) around the resonance.

On the other hand, if $\sigma_{VD} \! \gg \! \sigma_{ZD}$, at resonance involving dot $D$ we have $\sigma^2_{r,\pm} \! \approx \! \frac{1}{4}\sigma^2_{VD}$, 
It is important to note that far from the resonance, for $|\delta/2v|\! \ll \! 1$,
from the above Equations we see that the dephasing is visibly affected by valley splitting fluctuations as long as $|\delta_r/v_D| \! < \! \sqrt{\sigma_{VD}/\sigma_Z}$. This means that for large enough $\sigma_{VD}$ we can see strong effects of valley splitting fluctuations on dephasing time of $P_S(t)$ away from the near-resonance range of $B$ fields, in which the signal consists of three oscillations with appreciable amplitude. 

Finally, let us note that $\sigma^2_{r,0}$ goes to zero at resonance, as $\cos^2\theta_a \! \approx \! |\delta_r/2v_r|^2$ for $|\delta_r| \! \ll \! |v_r|$. Dephasing of this oscillation at resonance is then caused by second order terms in expansion of $\omega_0$ with respect to fluctuations. In order to simplify the discussion of this slow dephasing at such a ``sweet spot'' with respect to noise, let us assume that valley splitting fluctuations are much larger than eeman splitting fluctuations splitting fluctuations, so that close to resonance we have $\omega^0_r \! \approx \! 2|v_r| + \delta E_{V,r}^2/4|v_r|$.
 Instead of Gaussian decay, we obtain then an asymptotic power-law decay well known from calculations of dephasing due to quadratic influence of quasistatic noise \cite{Falci_PRL05,Ithier_PRB05,Koppens_PRL07,Ramon_PRB22}:
 \beq
\meanqs{\cos \omega^0_r t} = \frac{\cos(2|v_r|t + \frac{1}{2}\arctan t/\tau)}{(1+t^2/\tau^2)^{1/4}} \,\, , \label{eq:W0}
 \eeq
 from which we see that the half-decay time of envelope of oscillation is approximately $4\tau \! = \! 8|v_r|/\sigma^2_{V,r}$, and asymptotic decay is $\propto 1/\sqrt{t}$. As $8|v_r|/\sigma_{V,r}$ is typically large (see Section \ref{sec:dephasing_exp} for discussion involving realistic parameters of Si/SiGe quantum dots), this half-decay time is much larger than $T_{2,r,\pm}^* \! \approx \! 2\sqrt{2}/\sigma_{V,r}$ of the remaining oscillations close to the resonance. We can thus approximately treat the oscillation with $\omega^0_r \! \approx \! 2|v_r|$ frequency as nondecaying compared to the other two oscillations present in the signal close to the resonance.

\begin{figure}[tb]
\includegraphics[width=0.9\columnwidth]{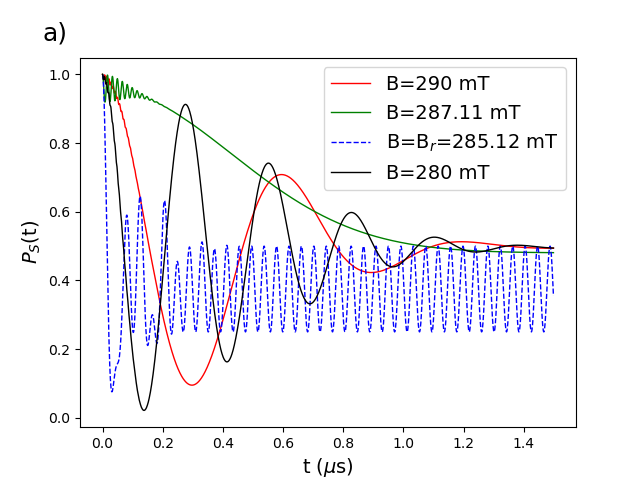}
\includegraphics[width=0.9\columnwidth]{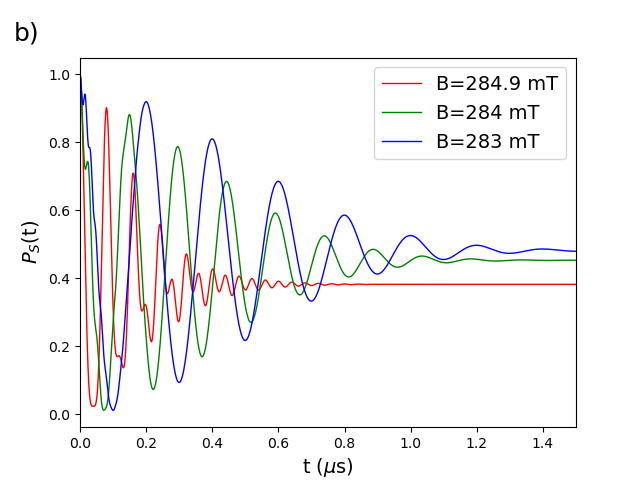}
    \caption{$P_{S}(t)$ signals calculated for resonance $r=1$ from  Table \ref{tab:resonances} calculated for parameters used in Fig.~\ref{fig:single}c: $\EVD{R}\! =\! 33$ $\mu$eV, $v_R \! =\! 50$ neV, $\Delta g \! =\! 6.58\cdot 10^{-4}$, $\sigma_{ZL}\! =\!\sigma_{ZR}\! =\! 1$ neV (corresponding to $T_{2}^{*}\! = \! 658$ for $B$ fields far from resonance), and $\sigma_{VR} \! = \! 10$ neV. In a) we show the results illustrating mostly the $B$-dependence of observed frequencies of oscillations: for $B=290$ mT and $280$ mT we see single oscillation (with frequency close to the unperturbed one given by $\Delta g \mu_B B/h \! \approx \! 2.6$ MHz in $B\! =\! 280$ mT case),  which decay due to fluctuations of spin splitting. At $B \! = \! 287.11$ mT the frequency of the oscillation with the largest amplitude is very close to zero, while at $B\! = B_r \! = \! 285.12$ mT we are exactly at the resonance: the two oscillations with larger amplitude exhibit beating and decay in $<300$ ns, and at longer times we see the non-decaying oscillation with frequency given by $2v_R/h$. In b) we show the evolution of the signal as the $B$ field is decreased from the resonance value: the red line, for $B$ field that differs from the resonance value $B_r$ by $\approx \! 0.2$ mT should be compared with the blue dashed line in (a) in order to see how narrow is the range of $B$ fields in which the oscillation with frequency $2v_R/\hbar$ is visible at long times. }
		\label{fig:PS_examples}
\end{figure}

An example dependence of the $P_{S}(t)$ on magnetic field in vicinity of a spin-valley resonance is shown in Fig.~\ref{fig:PS_examples}, where results of calculations using the above formulas for seven values of $B$ field in 10 mT wide range of fields around the resonance at $B_{r} \! \approx \! 285$ mT, corresponding to $E_{VR}\! =\! 33$ $\mu$eV. The calculations are performed with the parameters used previously in Fig.~\ref{fig:single}c for resonance in the $R$ dot, with $v_R\! =\! 50$ neV and $\sigma_{VR} \! =\! 10$ neV, while $\sigma_{ZL}\!=\!\sigma_{ZR} \! =\! 1$ neV. For $|B-B_r|\! \approx \! 5$ mT we see oscillations decaying with $T_{2}^{*} \! \approx \! \hbar/\sigma_{ZD}$, but with visibly distinct frequencies due to spin-valley mixing still having a non-negligible influence.
In Fig.~\ref{fig:PS_examples}a, at $B=287.11$ mT we have a result (green line) when the dominant frequency is approximately zero due to strong spin-valley induced renormalization. The quickly decaying low-amplitude component of this signal is due to a subdominant contribution that dephases on $\approx \hbar/\sigma_{VR}$ timescale. At $B\! =\! B_r$ (dashed blue line) we see fast dephasing of two dominant frequencies, with the sum of the two contributions to the signal exhibiting a beating pattern due to distinct values of these $\omega_r^{\pm}$, while for $t>200$ ns we see only the oscillation with $\omega_{r}^0 = 2|v_R|$ frequency that does not exhibit any decay on $\sim$ microsecond timescale. In Fig.~\ref{fig:PS_examples}b we see that after lowering the field by a fraction of a mT from $B_{r}$ (the red line), the oscillation with $\omega_{r}^0 = 2|v_R|$ frequency practically disappears due to reactivation of dephasing away from the ``sweet spot'', while the oscillations with $\omega_r^{\pm}$ are still undergoing dephasing in about 200 ns. As $B$ field is moved further away from the resonance (green and blue lines), the decay time of these oscillations grows towards the far-off resonance value of $T_{2}^{*} \! =\! 658$ ns.

\section{Comparison with experiments in Si/SiGe double quantum dots} \label{sec:experiment_comparison}

\subsection{Experiments on double quantum dots in two distinct Si/SiGe structures} \label{sec:exp}
\begin{figure*}[tb]
\includegraphics[width=0.9\textwidth]{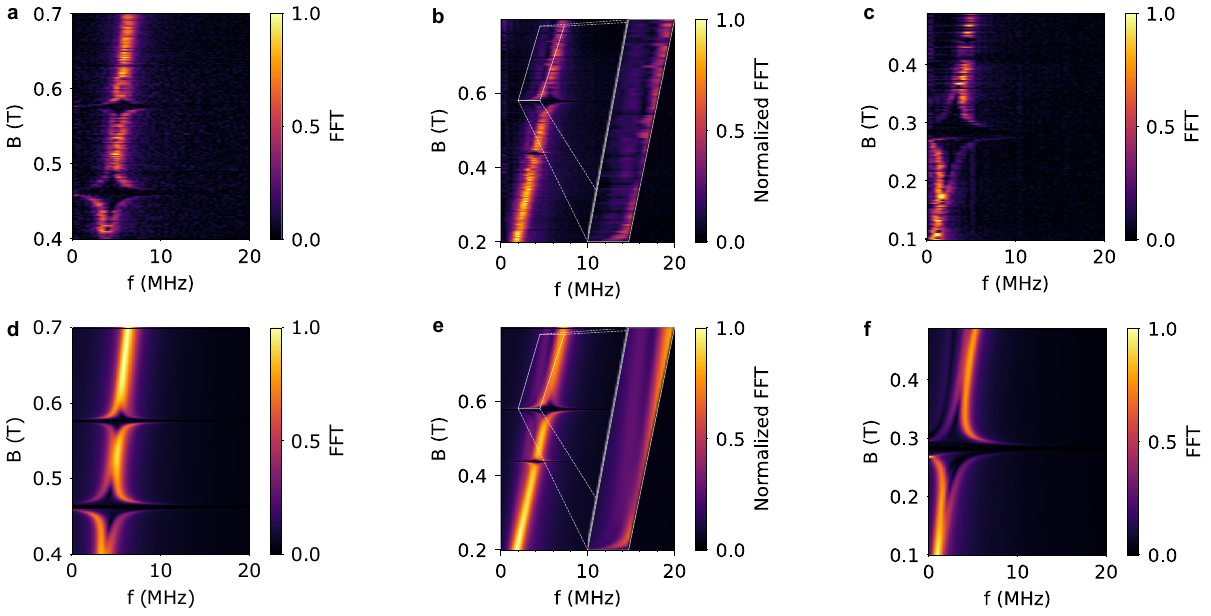}
\caption{(a)-(c): Fourier transform of singlet-return probability measured in a range of magnetic fields in structure from \cite{Struck_NC24,Volmer_NPJQI24} for three positions of the DQD along the $y$ axis (panels a,b,c correspond to datasets 1,2, and 3, respectively). (d)-(f): Theoretical fits to the data using parameters from Table \ref{tab:fit}. For dataset 2 in panels (b) and (e) we also show a zoom on an area in which the presence of frequency corresponding to $\Delta g_3$ in  Table \ref{tab:fit} is visible.}
\label{fig:exp123}
\end{figure*}

Charge separation from $(4,0)$ state resulting in initialization of $(3,1)$ spin singlet in a DQD, subsequent shuttling of the $R$ dot to location at distance $d$ from the initial one, followed by shuttling back to the tunnel-coupled DQD arrangement and Pauli spin blockade readout of singlet occupation, was studied in recent qubit shuttling experiments undoped Si/SiGe structures. 
In \cite{Struck_NC24,Volmer_NPJQI24} a structure having natural abundance of $^{29}$Si isotope was investigated, while in \cite{Volmer_arXiv25,Volmer_arXiv26} a structure containing isotopically enriched silicon with $<\! 800$ ppm of $^{29}$Si was studied. The structures came from completely distinct growth protocols, for details see \cite{Struck_NC24} and \cite{Volmer_arXiv25}, respectively.

We focus here on the experimental protocol in which the evolution of charge-separated singlet  occurs predominantly in the static two-dot configuration, with dot $L$ in its original place, and dot $R$ kept at distance $d$ from its original position for time $t$. This includes the case of $d\! =\! 0$, i.e.~of simply waiting in $(3,1)$ charge configuration for time $t$ and performing the PSB readout of singlet occupation. The experimental data presented in this Section corresponds to this case of no back-and forth shuttling of the $R$ dot. However, the previously described theory for singlet return probability signal applies also to the experiments with back-and-forth shuttling, provided that (1) shuttling to distance $d$ and back does not perturb the spin and valley state of the shuttled electron, and (2) the evolution in the relevant singlet-triplet subspaces during the motion of the $R$ dot is either negligible, or it has been accounted for by correcting the signal for the singlet-triplet oscillation that occurred during the motion of the $R$ dot. 
Let us however note that making sure  that assumption (1) holds requires caution. Simply shuttling as fast as possible with high-frequency oscillation of voltages on clavier gates \cite{Langrock_PRXQ23} in a conveyor belt setup can lead to changes of the valley state of the shuttled electron when conditions for adiabaticity of dynamics are broken at local minima of valley splitting \cite{Langrock_PRXQ23,DeSmet_NN25,Volmer_arXiv25}. 

First, let us discuss three sets of data for DQD with static $R$ dot acquired during the experiments described in \cite{Volmer_NPJQI24}. These sets were obtained for distinct values of screening gate voltages that resulted in shifting of the DQD position in the $y$ direction (in the plane of the QW and perpendicular to the line connecting $L$ and $R$ dot, which defines the $x$ axis that is in [110] or [1-10] direction). With the dataset 1 (D1) corresponding to a DQD at $y=0$,  the datasets 2 and 3 correspond to the DQDs at $y=-6$ and $6$ nm, respectively. 
The Fourier transform of the singlet return probability as function of frequency $f \! =\! \omega/2\pi$ and $B$ field is plotted for the three settings of screening voltages in Fig.~\ref{fig:exp123}(a,b,c). 
The frequency in $\tilde{P_S}(f,B)$ signal that is dominant in the whole range of $B$ fields was identified in \cite{Volmer_NPJQI24} for datasets shown in Fig.~\ref{fig:exp123}(a,b). For both, the $B$ dependence of this frequency was following the pattern from Fig.~\ref{fig:both}(b). Evolution of the observed pattern after shuttling of electron in $R$ dot allowed for identification of the resonance associated with the static dot, and consequently establishing that $\EVD{L}\! > \! \EVD{R}$ in the initial DQD.
 Using the definitions of $\DEZ{}$ and $\Delta g$ used here, from the qualitative shape of the $B$ dependence of the dominant frequency in the signal we arrive at the conclusion that $\Delta g\! <\! 0$ for the singlet contributing the largest-amplitude part of the signal (note that $\Delta g$ was defined in \cite{Volmer_NPJQI24} with an opposite sign, and positive value of $\Delta g$ given there is consistent with the discussion here).
The identification of the sign of $\DEZ{}$ was made in \cite{Volmer_NPJQI24} assuming that $\ket{S_{eg}}$ singlet was initialized. However, the analysis presented in Section \ref{sec:ST_oscillations} shows that this identification remains correct even if any other singlet was initialized.

\begin{table*}[t]
\begin{tabular}{|c|c|c|c|c|}
\hline	
													& D1 ($y=0$)  									& D2 ($y\!=\!-6$ nm)    							& D3  ($y\!=\!6$	nm)									& D4 (enriched Si sample) \\ \hline
$\EVD{L}$ ($\mu$eV) 			& $66.7\pm 0.07$  			& $66.7\pm 0.05$  			& --  									& $112.1\pm 0.007$  \\ \hline
$\EVD{R}$ ($\mu$eV) 			& $53.6 \pm 0.07$				& $50.7\pm 0.04$  			& $32.7\pm 0.06$  			& $95.53\pm 0.006$ \\ \hline
$v_L$ (neV) 							& $57.2 \pm 0.2$ 				& $68.5\pm 0.05$  			& --  									& $154.1\pm 0.05$   \\ \hline
$v_R$ (neV) 							& $90.5 \pm 0.2$				& $33.5\pm 0.06$  			& $136\pm 0.3$ 					& $43.96\pm 0.07$   \\ \hline
$\Delta g_1$ $(10^{-4})$	&  ${\bf -6.40\pm 0.04}$				& ${\bf -6.41\pm 0.02}$	& ${\bf -7.15\pm 0.1}$	& ${\bf 15.88\pm 0.02}$\\ \hline
$\Delta g_2$ $(10^{-4})$	& ${\bf 6.76\pm 0.08}$ 	&${\bf 6.68\pm 0.03}$		& ${\bf 6.9\pm 0.2}$		& ${\bf -15.6\pm 0.1}$\\ \hline
$\Delta g_3$ $(10^{-4})$	& $-7.03\pm 0.07$		& $5.22\pm 0.03	$				& $4.4\pm 1$						& $-33.93\pm 0.08$\\ \hline
$\Delta g_4$ $(10^{-4})$	& $7.39 \pm 0.07$				&$-4.95\pm 0.03$				& $-9.6\pm 0.9$					& $34.21\pm 0.09$ \\ \hline
$p_1$ $(10^{-2})$ 				& ${\bf 14.1 \pm 0.3}$	& ${\bf 14.92 \pm 0.03}$& ${\bf 12.2\pm 0.1}$		& ${\bf 7.83\pm 0.04}$\\ \hline
$p_2$ $(10^{-2})$					& ${\bf 8.0 \pm 0.3}$ 	& ${\bf 7.27\pm 0.03}$		& ${\bf 4.6\pm 0.1}$		& ${\bf 0.88\pm 0.02}$\\ \hline
$p_3$ $(10^{-2})$					& $4.04 \pm 0.3$ 				& $3.11\pm 0.03$				& $1.5\pm 0.1$					& $(-4.58\pm 2) \cdot 10^{-2}$\\ \hline
$p_4$ $(10^{-2})$ 				& $2.4 \pm 0.3$ 				& $2.38\pm 0.03$				& $-0.1\pm 0.9$					& $(-4.47\pm 2)\cdot 10^{-2}$\\ \hline
$\sigma_{Z}$ (neV) 				& $1.26 \pm 0.20$  			&  $1.395 \pm 0.055$  	&	$1.17\pm 0.19$  			& $0.48 \pm 0.72$  \\ \hline
$\sigma_{VL}$ (neV) 			& $5.72\pm 0.37$ 				&  $19.16 \pm 0.13$  		& --  									& $30.72 \pm 0.27$ \\ \hline
$\sigma_{VR}$ (neV) 			& $15.50\pm 0.43$  			&  $3.402 \pm 0.075$  	& $9.15 \pm 0.38$  			& $3.281 \pm 0.065$  \\ \hline
\end{tabular}
\caption{Parameters obtained by fitting the $P_S(t)$ signals the Fourier transforms of which are shown in Figs.~\ref{fig:exp123}(a,b,c) (respectively D1,D2,D3) and Fig.~\ref{fig:expnew}(a) (D4). $\EVD{D}$ are valley splittings in dot $D\! =\! L(R)$, $v_D$ are the spin-valley couplings in dot $D$, $\Delta g_k$ are the values of $g$-factor difference for up to four $(3,1)$ singlets that contribute to the signal, and $p_k$ are the amplitudes of the signals corresponding to these singlets. Note that negative $p_k$ means that the oscillation is phase-shifted by $\sim \pi$ with respect to the dominant part of the signal. $\sigma_Z$ is the fitted rms of spin splitting difference, and $\sigma_{VD}$ is the rms of valley splitting in dot $D$. 
For each dataset we have marked in bold the amplitudes of the two largest components of the signal, and the corresponding $\Delta g$. 
}
\label{tab:fit}
\end{table*} 

For data shown in Fig.~\ref{fig:exp123}c only one resonance is visible. Based on evolution of the pattern upon shuttling of $R$ dot we assign this resonance to this dot, and conclude that $\EVD{L}$ is either so large that the $L$ dot resonance is not visible in the measured range of fields (which was larger that the one in Fig.~\ref{fig:exp123}c, where we present a zoom on the single resonance), or $v_L$ was so small that the resonance could not be discerned. If $\EVD{L}\! > \! \EVD{R}$, we also have $\Delta g\! < \! 0$ also in this dataset.

Let us focus now on the features of the data that were not given attention in \cite{Volmer_NPJQI24}. A more careful look at the data in Fig.~\ref{fig:exp123}(a,b,c) shows that the observed patterns are in fact {\it sums of the two patterns from Fig.~\ref{fig:both}}, although the pattern from Fig.~\ref{fig:both}a has an amplitude smaller than the pattern from Fig.~\ref{fig:both}b by a factor of about 2. 
This is an unambiguous signature of the fact that {\it at least two $\ket{S_{\mu\nu}}$ singlets are initialized} by sweeping the detuning from $(4,0)$ to $(3,1)$ charge regime. As noted previously in Sec.~\ref{sec:nonadiabatic}, the presence of two components in the signal is not surprising, as it had been seen in a previous experiment on the same sample \cite{Struck_NC24} involving shuttling the $R$ dot for distance $d$ and immediately back. However, it is striking that the two strongest components in the measured signals correspond to very similar values of $|\Delta g|$: the two dominant parts of the signal that are clearly visible at resonances merge into a single line away from the resonances, showing that the values of $|\DEZ{}|$ for the two (or more) singlets are approximately equal to each other, with accuracy given by the typical linewidth of in Fig.~\ref{fig:exp123}, corresponding to $T_{2}^* \! \sim \! 1$ $\mu$s away from the resonances. 

In order to extract quantitative information from all the data we have performed fit to singlet return probability data obtained as function of time and magnetic field (with magnetic field range covering both the resonances, see Fig.~\ref{fig:exp123}), allowing for finite occupations $p_k$ of four singlets, ($k\! = \! 1 \ldots 4$) with corresponding values of $g$-factor difference $\Delta g_k$. For each dataset, we use the theoretical model described above that contains 15 parameters: $\EVD{D}$, $v_{D}$, $p_k$, $\Delta g_k$, $\sigma_{Z}$, 
and $\sigma_{VD}$. 
In the first round of fitting, we have concentrated on obtaining the values of parameters other than $\sigma_{ZD}$ and $\sigma_{VD}$, by using a simplified phenomenological model in which $T_{2}^{*}$ decay of all the contributions to $P_S(t)$ is allowed to vary in the vicinity of the resonances. Values of valley splitting and spin-valley couplings obtained from this fit are then used (as fixed) in the second round of fitting, in which we fit again $p_k$, $\Delta_k$, and also $\sigma_Z$ and $\sigma_{VD}$, while excluding the data very near the resonances (precisely for detuning $|\delta| \! \leq \! |v_D|$). The reason for this is twofold. Firstly, the presence of two oscillations with amplitudes of the same order of magnitude, but distinct frequencies, near the anticrossings due to opposite signs of respective $\Delta g$, leads to destructive interference effects in $P_S(t)$ signal that are hard to distinguish from effect of suppression of $T_{2}^*$ due to influence of valley splitting fluctuations. 
Secondly, a bit away from resonance, for $|\delta| \! > \! |v_D|$, we can use Eq.~(\ref{eq:sigma_off_even}) or (\ref{eq:sigma_off_odd}), respectively, depending on the dot in which the closest resonance occurs, to obtain a simple analytical model for $B$ field dependence of decay time of all the components of the signal. The results of the whole procedure are given in Table \ref{tab:fit}. Fourier transforms of $P_S(B;t)$ data calculated using these parameters are shown in the lower row of Fig.~\ref{fig:exp123}. 

The following features of the fitted sets of parameters are worth noting. For the three datasets, the ratios of maximal $p_1$ to the next largest $p_2$ are given by $1.75$, $2$, and $2.6$, respectively, while the relative differences of fitted $|\Delta g_1|$ and $|\Delta g_2|$ for the two components are between $3$\% and $6$\%. As one can see in the Figure, only in the second dataset (see the zoomed-in area in Fig.~\ref{fig:exp123}b) there is a reliable signature of the third component - in fact a pair components with $p_3 \! \approx \! p_4$ equal to about $0.3$ of $p_2$ - with $|\Delta g_3|\! \approx \! |\Delta g_4|$ being about $25\%$ smaller than $|\Delta g_{1(2)}|$. 
Note that components with finite $p_3$ and $p_4$ in dataset 1 are arguably spurious, and they simply contribute to additional line broadening of the signals associated with $p_1$ and $p_2$ amplitudes. 
Interestingly, the obtained magnitudes valley splitting fluctuations are larger by about an order of magnitude than the Zeeman splitting fluctuations, and they exhibit visible dot-dependence.


Let us stress that if the values of dot ($D)$ and valley ($\nu$) dependent $g$-factors, $g_{\nu_D}$, were independent random variables, a generic pattern arising from presence of two singlets should exhibit two discernible frequencies across the investigated range of magnetic fields, corresponding to distinct values of $|\Delta g_{\mu\nu}|$ and $|\Delta g_{\mu'\nu'}|$ for the $\ket{S_{\mu\nu}}$ and $\ket{S_{\mu'\nu'}}$ singlets present in the initialized mixed state. The results from Fig.~\ref{fig:exp123}(a,b,c) suggest a presence of nontrivial symmetry in the distribution of $g$-factors in the measured structure: the values of $\Delta g_{1}$ and $\Delta g_2$ extracted for the two highest-amplitude components of the signal have opposite signs, but nearly equal absolute values. 

\begin{figure}[tb!]
\includegraphics[width=0.9\columnwidth]{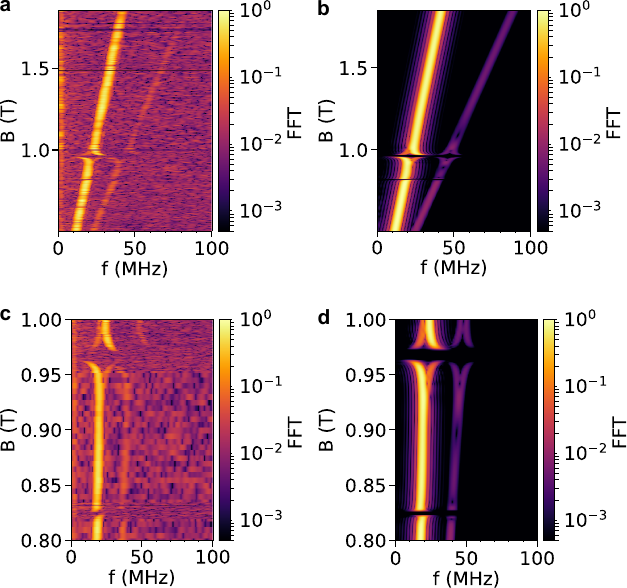}
\caption{Fourier transform of singlet return probability as function of frequency $f$ and magnetic field $B$: (a) experiment in isotopically enriched Si/SiGe double quantum dot and (b) theoretical fit with parameters given in the last column of Table \ref{tab:fit}. (c) and (d) are respective zoomed plots focused on the two visible resonances. Note that the color scale scale in this Figure is logarithmic. 
}
\label{fig:expnew}
\end{figure}

This hypothesis that such a symmetry is a general feature of Si/SiGe quantum dots is strengthened by an observation of analogous pattern of frequencies in DQD measurement in a distinct structure involving isotopically enriched Si used in shuttling experiments reported in \cite{Volmer_arXiv25}, see Fig.~\ref{fig:expnew}. Note that in this Figure the color scale is logarithmic, allowing one to see the presence of two distinct frequencies away from resonances, with one of the components having its amplitude larger by $\sim 100$ than the other. At the first sight one could conclude that this is precisely the above-mentioned ``generic'' pattern corresponding to a mixture of two initialized singlets, each having a random value of $\Delta g_{\mu\nu}$. However, a closer inspection of the results zoomed on the resonances again leads one to conclude that the signal corresponding to a single frequency away from the resonances is in fact a sum of two signals with $\Delta g_1 \! \approx \! -\Delta g_2$ (i.e.~the observed pattern being a sum of the two patterns shown in Fig.~\ref{fig:both}). We thus encounter again the presence of {\it two pairs of singlets, with opposite signs but equal moduli of $\Delta g$ within pair}. 
The data shown in Fig.~\ref{fig:expnew}a can be fitted by a sum of four contributions, a dominant one corresponding to $\Delta g \! \approx \! 15.9 \cdot 10^{-4}$,  the subdominant one (with amplitude smaller by a factor of $\sim  \! 10$) with $\Delta g \! \approx -15.6\cdot 10^{-4}$, and two much weaker components (with amplitudes smaller by one more order of magnitude) with $\Delta g \! \approx \! \pm 34\cdot 10^{-4}$, see Table \ref{tab:fit}. Note that the values of $\Delta g$ in this sample are larger by a factor of at least 2 that in the previously discussed one (this is confirmed by comparison of  measurements of $g$-factor maps spanning an area much larger than the sizes of the quantum dots in this sample \cite{Volmer_arXiv26} and in the one studied in \cite{Volmer_NPJQI24}). The ratio $p_1/p_2\! \approx \! 10$ is much larger that in the other structure, and as noted before $p_1/p_{3(4)} \! \approx \! 100$. This shows that the effects of non-adiabaticity during charge separation are weaker in this structure, suggesting larger values of $t_{\mu\nu}$ - although the presence of $4$ components in the fit to the signal shows that all the valley occupations of $(3,1)$ singlets are still initialized with finite probability.

\subsection{Dephasing seen in experiments} \label{sec:dephasing_exp}
Let us discuss now the values of standard deviations of fluctuations of spin and valley splitting. 
The value of $\sigma_Z$ in this isotopically enriched sample is smaller by at least a factor of $2$ than in the sample with natural silicon. Qualitatively this is expected. Quantitative discussion of the obtained values is left for future research. 

Valley splitting fluctuations are caused by electric field noise: 	fluctuations in $z$ component of electric field modify the overlap of the electron wavefunction with Si/SiGe interface, and fluctuations of the in-plane components of the field shift the position of the dot. Both effects should lead to changes of valley splitting for a given quantum dot. We focus here on the latter effect, as the results from Table I, and the valley splitting mappings results from \cite{Volmer_NPJQI24} provide us with estimates of derivatives of $\EVD{}$ with respect to in-plane coordinates. Traces of $\EVD{R}(x)$ from \cite{Volmer_NPJQI24} imply $|\mathrm{d}\EVD{R}/\mathrm{d}x| \! \approx \! 0.5$ $\mu$eV/nm  for DQDs from Fig.~\ref{fig:exp123}. On the other hand, from values given in Table \ref{tab:fit}, recalling that DQDs in D2 and D3 are shifted by about $\pm 6$ nm along $y$ axis with respect to the DQD corresponding to D1 set of results, we get two values of $|\mathrm{d}\EVD{R}/\mathrm{d}x| \! \approx \! 0.5$ $\mu$eV/nm and $\approx \! 3$ $\mu$eV/nm. We can then estimate the order-of-magnitude of in-plane fluctuations of QD position from experiments on spin decoherence caused by electric field noise in presence of magnetic field gradient. From \cite{Struck_NPJQI20} one can extract the standard deviation of QD position due to $1/f$ charge noise (with data acquisition time of about 10 minutes, which is similar to the time used for each trace in the discussed experiments) of $\sigma_x \! \approx \! 5\cdot 10^{-3}$ nm. A similar value can be inferred from levels of in-plane electric field noise measured in \cite{Yoneda_NP23}. Using this value of $\sigma_x$, we arrive at $\sigma_V \! \approx \! 5$ neV for $|\mathrm{d}\EVD{D}/\mathrm{d}x| \! =\! 1$ $\mu$eV/nm. Taking into account that we have used the estimates of $\sigma_x$ from other experiments on distinct Si/SiGe structures, so that it should be treated as an order-of-magnitude estimate, we can thus state that values of $\sigma_V$ of a few neV are expected, and $\sigma_V \sim 10$ neV is consistent with mechanism of charge noise induced in-plane motion in presence of measured spatial dependence of valley splitting.

\subsection{Implications for valley-dependence of $g$-factors}  \label{sec:g_factor}
These observations concerning the pattern of $g$-factor differences in the measured DQDs are in fact naturally explainable by the recently developed theoretical model of valley dependence of $g$-factors in Si/SiGe quantum dots \cite{Woods_arXiv24,Woods_arXiv25}, which predicts 
\begin{equation}
g_{\nu_D} \approx g_0 \pm \delta g_0 \cos\phi_D \equiv g_0 \pm \delta g_{D} \,\, , \label{eq:g}
\end{equation} 
in which $+$ ($-$) sign corresponds to $\nu \! =\! e$ ($g$) valley state, $g_0$ and $\delta g_{0}$ are weakly dependent on dot position and shape (assuming that the dot shape does not change drastically when it is moved from one place to another in a given heterostructure), and $\phi_D$ is the phase of the valley coupling matrix element for electron in dot $D$, $\Delta_D \! \equiv\! |\Delta_D| e^{i\phi_D}$ \cite{Losert_PRB23}. According to recent theoretical works \cite{Wuetz_NC22,Losert_PRB23,Lima_MQT23,Thayil_PRB25} discussing the influence of alloy disorder on the value of $\Delta(\mathbf{r})$ for quantum dot centered at position $\mathbf{r}$, real and imaginary components of $\mathbf{r}$ are Gaussian random fields with correlation length of the order of quantum dot size. 

The opposite signs of valley-dependent correction to $g_0$ for the two valleys immediately give us that for singlets created by intra-valley (i.e.~valley eigenstate conserving) tunneling, $\ket{S_{gg}}$ and $\ket{S_{ee}}$, we have 
\begin{equation}
\DEZ{ee} = \delta g_L-\delta g_R = -\DEZ{gg} \,\, ,
\end{equation}
and for the singlets created by inter-valley tunneling we obtain an analogous relationship between the $g$-factor differences
\begin{equation}
\DEZ{eg} = \delta g_L+\delta g_R = -\DEZ{ge} \,\, .
\end{equation}
If the inter-dot tunneling is predominantly of the valley-conserving character, i.e.~$|t_{ee}|\! =\! |t_{gg}| \! \gg \! |t_{eg}|\!=\! |t_{ge}|$, see Eq.~(\ref{eq:tmunu}), so that $S_{ee}$ and $S_{gg}$ singlets have a chance to be created by adiabatic passage, with both $p_{ee}$ and $p_{gg}$ both non-negligible, while the tunnel anticrossings involving $S_{eg}$ and $S_{ge}$ singlets are passed almost diabatically due to smallness of the valley-flip tunneling, we expect the $P_S (t)$ signal to contain two components corresponding to very similar values of $\Delta g$, but opposite signs - just as in the results shown in Figs.~\ref{fig:exp123} and \ref{fig:expnew}. The same is expected to occur when the valley-flip tunneling is much stronger than the valley-conserving tunneling. 

All the experimental results discussed above can be explained in a natural way using this model. The results for DQDs with non-shuttled $R$ dot in structure from \cite{Volmer_NPJQI24} correspond to either $|\delta g_L-\delta g_R|$ or $|\delta g_L + \delta g_R|$ being equal to approximately $7 \cdot 10^{-4}$, with little variation in $\Delta g$ exhibited when the DQD is moved in the $y$ direction by a distance of 6 nm. The appearance of component with distinct $|\Delta g| \! \approx \! 5 \cdot 10^{-4}$ in the second dataset can be then explained by assuming that the value of $|\delta g|$ in one dot is much larger than in the other: precisely, assumption of $|\delta g| \! \approx \! 5.75 \cdot 10^{-4}$ in one dot, and $|\delta g| \! \approx \! 0.75 \cdot 10^{-4}$ in the other, reproduces the observed set of $\Delta g$ values. On the other hand, if we approximate the fit to the fourth dataset by $|\Delta g_{1(2)}| \! \approx \! 16 \cdot 10^{-4}$ and $|\Delta g_{1(2)}| \! \approx \! 34 \cdot 10^{-4}$, we arrive at $|\delta g| \! \approx 24 \cdot 10^{-4}$ in one dot, and $|\delta g| \! \approx \! 9 \cdot 10^{-4}$ in the other. Note that depending on assignment of $ee/gg$ and $eg/ge$ valley character to each pair of singlets, and on the choice of assignment of signs of $\Delta g$ within each pair, any of 16 possible assigments of 
$(\pm 9 \cdot 10^{-4}, \pm 25 \cdot 10^{-4})$ to $(\delta g_L, \delta g_R)$ can be realized.

\section{Summary }
We have presented a theoretical model of initialization, dynamics, and readout of $(3,1)$ singlets in silicon double quantum dots in which spin-valley coupling is non-negligible. We have given a detailed account of the influence of spin-valley coupling on dynamics of these singlets, providing a full picture of the effect of large spin-valley coupling induced renormalization of frequency of singlet return probability signal.
 The presence of this renormalization was used in \cite{Volmer_NPJQI24, Volmer_arXiv25}, and more recently in \cite{Volmer_arXiv26}, to perform the mapping of valley splitting by shuttling one of the dots from the standard DQD setup (the two dots $\leq 100$ nm away, with tunnel coupling between them controlled by the barrier gate) by up to a few hundreds of nanometers from its initial position. 
 
Comparison of the predictions of the model with the experiments conducted on two distinct Si/SiGe structures has shown the following. (1) charge separation from $(4,0)$ to $(3,1)$ cannot be assumed to be perfectly adiabatic, and initialization of the statistical mixture of $(3,1)$ singlets with distinct valley occupation patterns occurred in both experiments. (2) While creation of the mixture of valley occupations was not intended, careful analysis of the singlet return probability signals near the spin-valley resonances has uncovered an interesting symmetry of the $g$-factors in the studied dots. These results give support to the recently proposed theory of valley-dependence of $g$-factors in Si/SiGe quantum dots \cite{Woods_arXiv24,Woods_arXiv25} (note that the same symmetry was predicted for SiMOS dots with large valley splittings in \cite{Ruskov_PRB18}). (3) Valley splitting fluctuations were predicted to contribute to dephasing of singlet-triplet oscillations in the vicinity of the spin-valley resonances, and to dominate this dephasing if these fluctuations are larger than the spin splitting fluctuations (due to nuclear noise of $g$-factor variations caused by electric field noise). Fits of the model to the data from two distinct structures have given strong hints that the dephasing time indeed decreases by at least an order of magnitude in the vicinity of the resonances. 

Let us stress that while signatures of occupation of multiple valley states after charge separation were observed before \cite{Struck_NC24}, here they have been analyzed in more detail, and a basic model for the influence that nonadiabatic valley dynamics during detuning sweeps has on singlet initialization and Pauli Spin Blockade measurement was analyzed in Sec.~\ref{sec:nonadiabatic}. Insight given there should inform future work on optimization of fidelity of spin qubit initiation and readout when valley splittings in the structure are not very large, and spatial inhomogeneity of valley couplings leads to finite tunnel couplings between all the valley states in the two adjacet quantum dots.

\section*{Data availability}
The data that support the findings of this study are available in the \href{https://zenodo.org/records/19605349?token=eyJhbGciOiJIUzUxMiJ9.eyJpZCI6ImEyY2JjOGQ5LWY1MmItNDlmNi1hNGIwLTBkNjYxNmQxNDJhZiIsImRhdGEiOnt9LCJyYW5kb20iOiJkNWVlNjk3ODUxYjg1Yjg0NGE4ZDBiMTcwOWU2NDY1MSJ9.OJdxzNd5JWWzA9Db_ymqEmskDTn7_pJ78Li2mXLmRahuIFDFGxMvZ6BGSj2SZq_IHeQukyMTzlDXtImCdCV65g}{Zenodo repository}.

\acknowledgements
We acknowledge support of the Dresden High Magnetic Field Laboratory (HLD) at the Helmholtz-Zentrum Dresden - Rossendorf (HZDR), member of the European Magnetic Field Laboratory (EMFL). This work was funded by the German Research Foundation (DFG) within the project 289786932 (SCHR 1404/2-2) and under Germany's Excellence Strategy - Cluster of Excellence Matter and Light for Quantum Computing" (ML4Q) EXC 2004/2 - 390534769, and  by the European Union’s Horizon 2020 Research and Innovation Actions under Grant Agreement No. 101174557 (QLSI2).
The device fabrication has been done at HNF - Helmholtz Nano Facility, Research Center Juelich GmbH \cite{Albrecht17}.

%

\end{document}